\documentclass[article, superscriptaddress]{revtex4}

\usepackage{graphicx,bm,epsfig}
\usepackage[a4paper,top=3cm,bottom=2.7cm,left=2.5cm,right=2.5cm]{geometry}

\def\spose#1{\hbox to 0pt{#1\hss}}
\def\lesssim{\mathrel{\spose{\lower 3pt\hbox{$\mathchar"218$}}
 \raise 2.0pt\hbox{$\mathchar"13C$}}}
\def\gtrsim{\mathrel{\spose{\lower 3pt\hbox{$\mathchar"218$}}
 \raise 2.0pt\hbox{$\mathchar"13E$}}}

\def\<{\langle}
\def\>{\rangle}

\begin{document}

\title{Integral-equation analysis of single-site coarse-grained models 
for polymer-colloid mixtures}

\author{Roberto Menichetti}
\email{Roberto.Menichetti@roma1.infn.it}
\author{Andrea Pelissetto}
\email{Andrea.Pelissetto@roma1.infn.it}
\affiliation{Dipartimento di Fisica, Sapienza Universit\`a di Roma, P.le Aldo Moro 2, I-00185 Roma, Italy.}
\affiliation{INFN, Sezione di Roma I, P.le Aldo Moro 2, I-00185 Roma, Italy.}
\author{Giuseppe D'Adamo}
\email{giuseppe.dadamo@sissa.it}
\affiliation{SISSA, V. Bonomea 265, I-34136 Trieste, Italy.} 
\author{Carlo Pierleoni}
\email{Carlo.Pierleoni@aquila.infn.it}
\affiliation{Dipartimento di Scienze Fisiche e Chimiche, Universit\`a dell'Aquila and CNISM, UdR dell'Aquila, V. Vetoio 10, Loc. Coppito, I-67100  L'Aquila, Italy.}

\begin{abstract}
We discuss the reliability of integral-equation methods based on 
several commonly used closure relations in determining the phase diagram
of coarse-grained models of soft-matter systems characterized 
by mutually interacting soft and hard-core particles. Specifically, 
we consider a set of potentials 
appropriate to describe a system of hard-sphere colloids 
and linear homopolymers in good solvent, and investigate 
the behavior when the soft particles are smaller than 
the colloids, which is 
the regime of validity of the coarse-grained models. 
Using computer-simulation results
as a benchmark, we find that the hypernetted-chain approximation provides 
accurate estimates of thermodynamics and structure in the colloid-gas 
phase in which the density of colloids is small. On the other hand, 
all closures considered appear to be unable to describe the behavior of the 
mixture in the colloid-liquid phase, as they cease to converge at 
polymer densities significantly smaller than those at 
the binodal. As a consequence, integral
equations appear to be unable to predict a quantitatively
correct phase diagram.
\end{abstract}

\maketitle


\section{Introduction}

Integral-equation methods are a very powerful tool to 
determine the thermodynamics and the liquid structure of simple fluids
\cite{HansenMcDonald,Attard-02}. They rely on different approximate
closure relations which, supplemented by the 
Ornstein-Zernike (OZ) equation, allow a direct and numerically fast 
determination of the pair correlation functions as well as of 
thermodynamic quantities like pressure, 
compressibility, chemical potential, $\ldots$ 
For simple fluids these methods cannot compete nowadays with 
Monte Carlo and molecular-dynamics simulations. Nonetheless, they have the 
advantage of providing reasonably accurate estimates of 
thermodynamic quantities with a very limited effort, and they are 
therefore a very valuable tool when the system under investigation 
depends on many parameters,
for instance in the case of multicomponent systems. Moreover, they are 
still very useful for the analysis of systems for which 
atomistic simulations are particularly slow, for instance in glassy systems;
see, e.g., Refs.~\cite{MP-99,PZ-10,BHP-14}. 

Liquid-state integral equations have also been extensively used to compute 
fluid-fluid phase-coexistence lines. In the density region in which 
the system demixes, integral equations may not converge, or may converge
to physically unacceptable solutions. The relation between the boundary
of this nonconvergence region (we will call it termination line)
and the binodal and the spinodal curves 
characterizing the two-phase unstable region
has been the subject of many studies, 
see, e.g., Refs.~\cite{CS-83,Belloni-93,RVL-96,SL-05}. 
In particular, it has been shown that, except in the case of very 
simple approximations, thermodynamical quantities 
do not show any particular divergence on this line, hence it 
cannot be taken as an approximate estimate of
the spinodal line. However, it is usually assumed that it is somewhat close to 
the line where phase separation occurs.

In this paper we wish to investigate the reliability of integral-equation
methods for the determination of the phase diagrams of typical 
coarse-grained models of soft-matter systems. We
consider here a binary mixture of soft and hard spheres 
of different sizes with an intrinsic nonadditive nature. 
Although we take specific pair
potentials, appropriate to describe, in a coarse-grained fashion, 
a binary system of hard-sphere colloids and long polymers under good-solvent
conditions \cite{Likos-01,HL-02}, the conclusions 
should apply to a general class of soft-matter systems 
that can be modelled as mixtures of soft and/or hard spheres,
interacting via short-range potentials
\cite{LFH-00,DLL-02,FHL-03,PSCG-06,PP-14}.
The phase diagram of the 
coarse-grained model has been accurately determined in 
Ref.~\cite{DMPP-15}, by means of Monte Carlo simulations, for different
values of the polymer-to-colloid size ratio. 
Here, we investigate the same problem by using 
integral-equation methods. We employ the hypernetted-chain (HNC) ,
the Percus-Yevick (PY), the Rogers-Young (RY), and the reference HNC (RHNC)
closures \cite{HansenMcDonald,RY-84,RA-79,ELLAL-84}. For each of them
we determine the termination line, whose position is then compared 
with the Monte Carlo binodal with the purpose of understanding if 
this line provides a reasonable approximation of the boundary of the 
two-phase region. For small  polymer densities, we will also be able to 
compute by Monte Carlo simulations the bridge 
functions---quantities that have an intrinsic interest 
in liquid-state theories---which can then be 
compared with the approximate ones considered in the different approaches.

The paper is organized as follows.
In Sec.~\ref{sec2} we define the model,
report the definitions of the different closures
we use, and the explicit expressions of the quantities that are 
considered in the paper.
In Sec.~\ref{sec3} we present our results.
In Sec.~\ref{sec3.1} we determine the termination line for the different 
closures for two different values of the polymer-to-colloid 
size ratio $q$, $q = 0.5$ and $q=0.8$. In Sec.~\ref{sec3.2} 
we compare the integral-equation 
predictions for structure and thermodynamics with Monte Carlo results.
In Sec.~\ref{sec3.3} we determine the bridge functions with
Monte Carlo methods and compare them with 
those used in the different integral-equation approaches.
In Sec.~\ref{sec3.4} we consider a novel approximation that uses the 
Monte-Carlo determined bridge functions. Finally, in Sec.~\ref{sec4} we 
draw our conclusions. Technical details are reported in Appendix 
\ref{AppA}.
The explicit expressions of the potentials are reported in 
Appendix \ref{AppB}.

\section{Definitions} \label{sec2}

\subsection{The model} \label{sec2.1}

\begin{figure}[b]
\begin{center}
\begin{tabular}{cc}
\epsfig{file=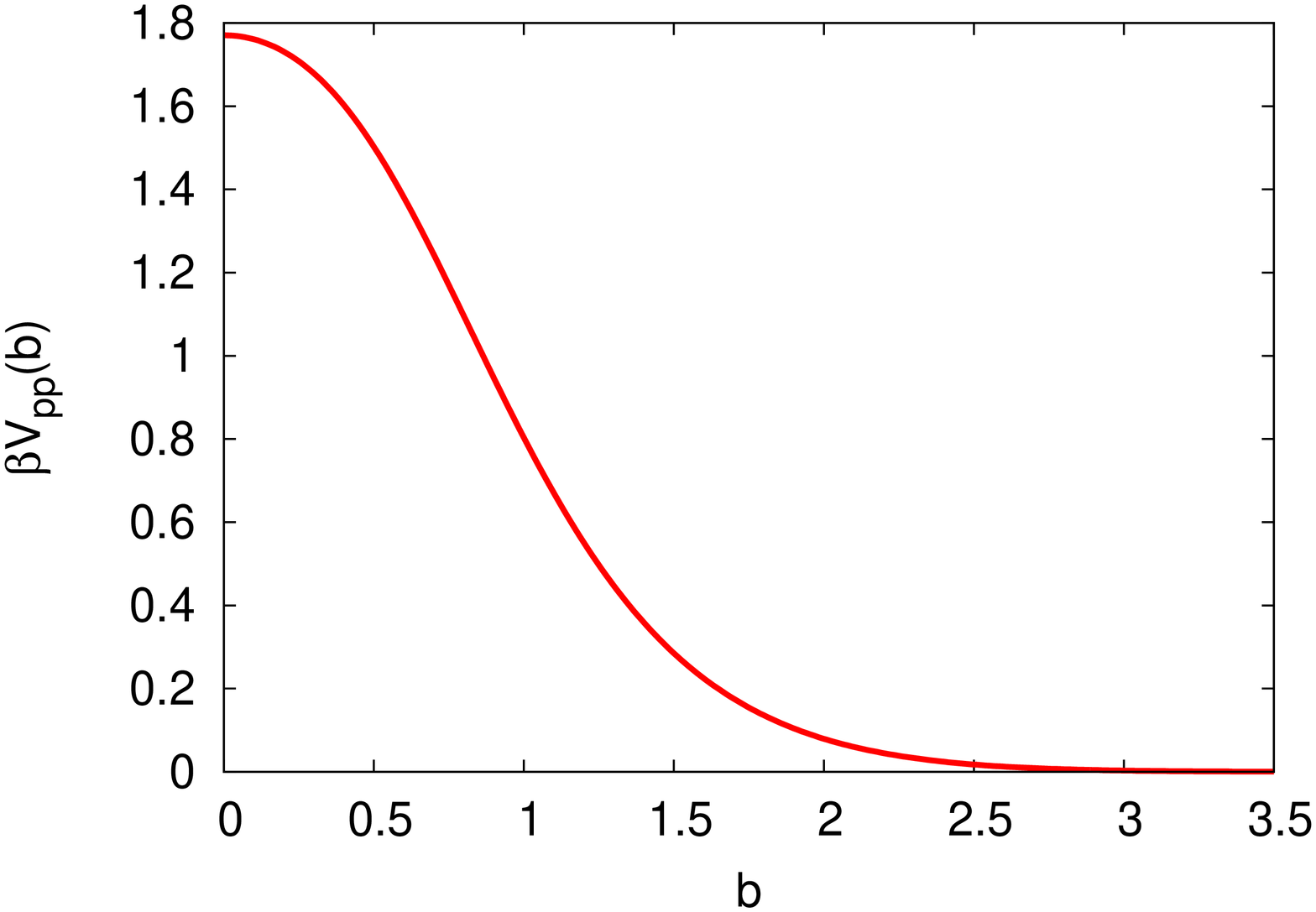,width=4.8truecm,angle=-90} 
   \hspace{0truecm} &
\epsfig{file=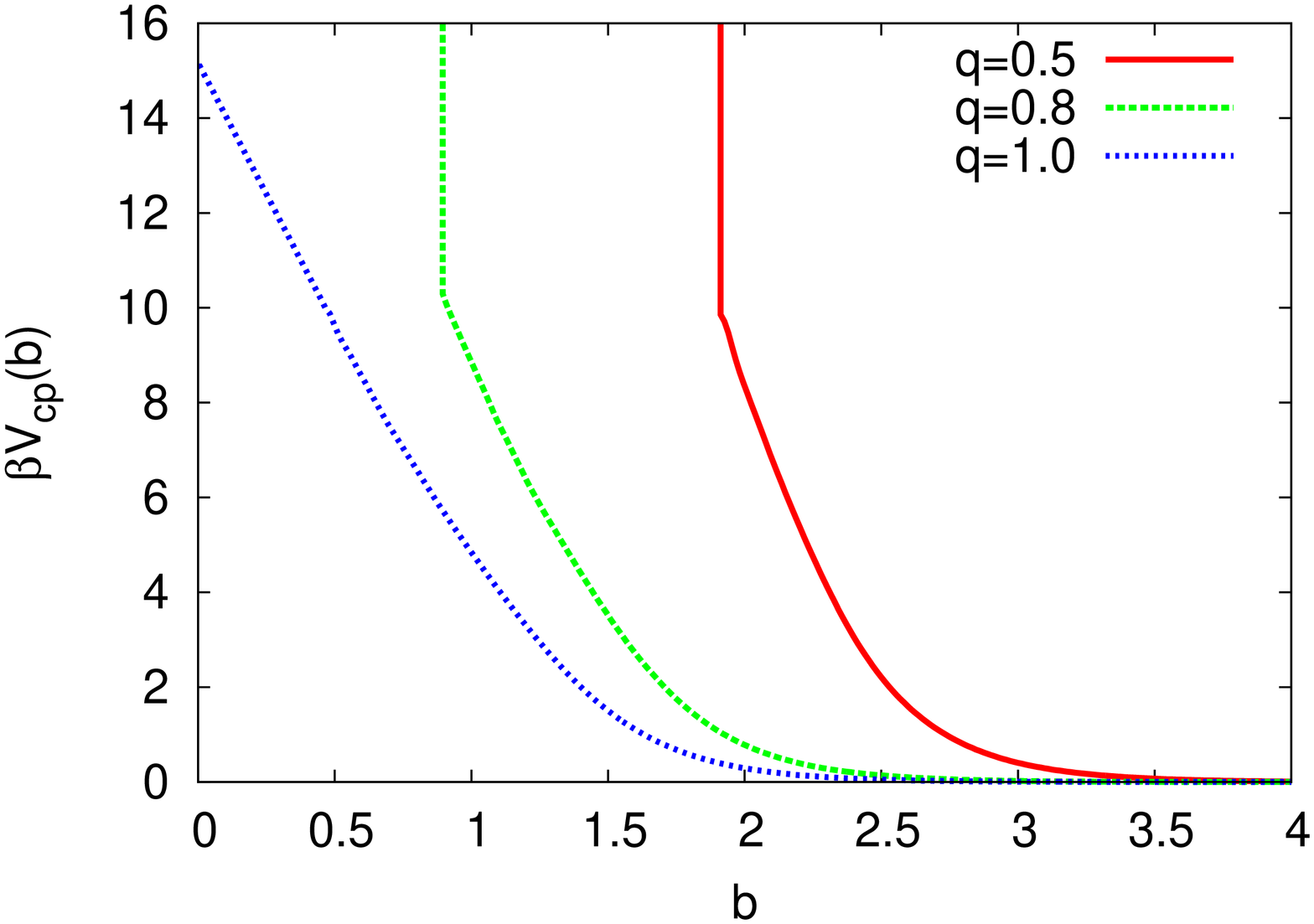,width=4.8truecm,angle=-90} 
\hspace{0truecm} \\
\end{tabular}
\end{center}
\caption{Left: polymer-polymer pair potential $\beta V_{pp}(b)$
as a function of $b = r/R_g$; right: polymer-colloid 
potential $\beta V_{cp}(b;q)$ as a function of $b = r/R_g$
for $q = 0.5,0.8,1$. For $q = 0.8$ and $q = 0.5$, we assume
$V_{cp}(b;q) = \infty$ for $ b < 0.90$, 1.91, respectively.
}
\label{pair-potentials}
\end{figure}

We consider a mixture of mutually interacting 
hard spheres  of radius $R_c$ and of 
soft particles with a typical interaction range $R_g$. 
Specifically, we consider here a set of potentials which are
appropriate to describe a system of hard-sphere colloids of 
size $R_c$ and linear homopolymers in good solvent of radius of 
gyration $R_g$, after tracing out the monomer degrees of freedom and 
replacing each chain with a particle coinciding with its center of mass.
The coarse-grained model is accurate only if polymers are dilute,
i.e., for $\Phi_p = 4 \pi N_p R^3_g/(3 V)\lesssim 1$
($N_p$ is the number of colloids in the volume $V$),
and if the polymer-to-colloid size ratio $q=R_g/R_c$
satisfies $q\lesssim 1$ \cite{DPP-14-colloids,DMPP-15}
(a discussion of the accuracy of 
the model, with a comparison with full-monomer results is presented
in Ref.~\cite{DMPP-15}). 

Polymer-colloid solutions have been 
extensively studied \cite{Poon-02,FS-02,TRK-03,MvDE-07,FT-08,ME-09},
because of their rich phase diagram, which presents fluid-fluid and fluid-solid 
coexistence lines, and because of their technological relevance \cite{LT-11}.
In this paper, we will not be interested in using the model to predict 
their phase behavior. Rather, we take it as a
typical soft-matter system and use it as reference 
model for which we can study the predictivity of the 
different closures that are typically used in integral-equation studies.
We will consider three different values of $q = R_g/R_c$, $q = 0.5, 0.8$, and 1.
The corresponding pair potentials have been determined in several papers 
\cite{LBHM-00,LBMH-02-a,LBMH-02-b,PH-05,PH-06,DPP-13-depletion}. Here
we shall use the accurate scaling-limit results of
Refs.~\cite{PH-05,DPP-13-depletion}. 
They are reported for completeness 
in Appendix \ref{AppB} and shown in Fig.~\ref{pair-potentials}.
The polymer-polymer potential is essentially Gaussian with 
$V_{pp}(b=0) \approx 1.8 k_B T$ at full overlap. On the other hand, 
the nature of the polymer-colloid potential depends on $q$.
For $q\le 1$ the potential is expected to be 
infinite at full overlap and very large for 
$r \lesssim R_c = R_g/q$. Then, it decays fast, with a tail that is small 
for $r \gtrsim R_c + 2 R_g$. Note that we have not been able to 
determine $\beta V_{cp}(b;q)$ in the small-$r$ region in which 
$\beta V_{cp}(b;q) \gtrsim 10$. For these values of $r$ we simply assume
$\beta V_{cp}(b;q) = +\infty$ for $q = 0.5$ and $q = 0.8$. For $q = 1$ 
we performed a linear extrapolation.

\subsection{Closure relations} \label{sec2.2}

In the integral-equation approach the basic ingredients are the 
pair correlation functions
$h_{\alpha\beta}(r)$ ($\alpha$ and $\beta$ label the two species) and the 
direct correlation functions $c_{\alpha\beta}(r)$. They are related by
the Ornstein-Zernike (OZ) \cite{HansenMcDonald} relations
\begin{equation}
\hat{h}_{\alpha\beta} (k) = \hat{c}_{\alpha\beta}(k) + 
   \sum_\gamma \hat{c}_{\alpha\gamma}(k) \rho_\gamma \hat{h}_{\gamma\beta}(k),
\end{equation}
where we denote with $\hat{f}(k)$ the (three-dimensional) Fourier transform 
of any function $f(r)$. To compute the quantities of interest, 
the OZ relation must be 
supplemented by a closure relation, which can be written in general form as 
\begin{equation}
g_{\alpha\beta}(r) = e^{-\beta V_{\alpha\beta}(r)} \exp[h_{\alpha\beta}(r) - 
c_{\alpha\beta}(r) + b_{\alpha\beta}(r)],
\label{def-bridge}
\end{equation}
where $V_{\alpha\beta}(r)$ are the pair potentials,
$g_{\alpha\beta}(r) = h_{\alpha\beta}(r) + 1$ is the pair distribution 
function and $b_{\alpha\beta}(r)$ is the so-called bridge function. The latter
quantity cannot be computed exactly, hence we consider several different 
approximations:
\begin{itemize}
\item[(i)] 
Hypernetted chain (HNC) closure 
\cite{HansenMcDonald}. We simply set $b_{\alpha\beta}(r) = 0$
for all $\alpha,\beta$. 
This approximation is very accurate for soft potentials
\cite{HansenMcDonald}.
\item[(ii)] 
Mixed HNC/Percus-Yevick (PY) closure. For hard spheres the PY closure relation 
\cite{HansenMcDonald}
\begin{equation}
g_{\alpha\beta}(r) = e^{-\beta V_{\alpha\beta}(r)} [1 + h_{\alpha\beta}(r) - 
c_{\alpha\beta}(r)]
\label{PY-closure}
\end{equation}
is more accurate than the HNC closure. Here we consider the HNC 
closure for polymer-polymer and polymer-colloid correlations and the PY closure
for the colloid-colloid correlations. 
\item[(iii)]
Rogers-Young (RY) closure 
\cite{RY-84,BH-91}. 
This closure mixes the HNC and the PY closures, adding free parameters
that are tuned to obtain thermodynamic consistency. It is defined by
\begin{equation}
g_{\alpha\beta}(r) = e^{-\beta V_{\alpha\beta}(r)} \left[
 1 + {\exp[(h_{\alpha\beta}(r) - c_{\alpha\beta}(r)) f_{\alpha\beta}(r)] - 1 
      \over f_{\alpha\beta}(r) } \right],
\end{equation}
where the function $f_{\alpha\beta}(r)$ is given by 
\begin{equation}
   f_{\alpha\beta} = 1 - e^{-\chi_{\alpha\beta} r}.
\end{equation}
Note that, for $\chi_{\alpha\beta} \to 0$ we recover the PY closure, while 
in the opposite limit, $\chi_{\alpha\beta} \to \infty$, 
we reobtain the HNC closure.
In most of the discussion we have considered a single optimization parameter,
setting $\chi_{\alpha\beta} = \chi/s_{\alpha\beta}$, 
$s_{cc} = R_c$, $s_{pp} = {R}_g$, $s_{pc} = (R_c + {R}_g)/2$. The parameter
$\chi$ has been determined as discussed below.
\item[(iv)] 
Reference HNC (RHNC) closure \cite{RA-79,ELLAL-84}. In this approach
one sets $b_{\alpha\beta}(r) = b^{HS}_{\alpha\beta}(r;R_p,R_c)$, where the 
latter quantities are the bridge functions of a system of additive hard spheres
of radii $R_p$ and $R_c$ at the same densities of the polymers
and colloids in the original system.
The polymer effective radius $R_p$ is determined by using 
the Lado criterion \cite{Lado-82,ELLAL-84}
\begin{equation}
\sum_{\alpha\beta} x_\alpha x_\beta
   \int_0^\infty r^2 dr\, 
  [h_{\alpha\beta}(r) - h_{\alpha\beta}^{HS}(r;R_p,R_c)] 
  {\partial b^{HS}_{\alpha\beta}(r;R_p,R_c) \over \partial R_p} = 0,
\label{Lado-eq}
\end{equation}
where $x_\alpha = N_\alpha/(N_c + N_p) = \rho_\alpha/(\rho_p + \rho_c)$.
The bridge functions $b^{HS}_{\alpha\beta}(r;R_p,R_c)$ can be computed 
as discussed in Refs.~\cite{Lebowitz-64,LHP-65,MCSL-71,VW-72,GH-72,LL-73,
HG-75,ELLAL-84}.
\end{itemize}
Solving simultaneously the OZ and the closure relations, one obtains $h_{\alpha\beta}(r)$
and $c_{\alpha\beta}(r)$. Then, one can use them to compute thermodynamic quantities.

\subsection{Observables} \label{sec2.3}

We will be interested in computing the pressure. One possibility 
consists in using the virial expression:
\begin{equation}
\beta P^{\rm (vir)} = \rho \left(1 + \sum_{\alpha\beta} Z_{\alpha\beta}\right),
\end{equation}
where $\beta = 1/k_B T$,
$\rho = \rho_p + \rho_c$, and the quantities $Z_{\alpha\beta}$ are given by
\begin{equation}
Z_{\alpha\beta} = - {2\pi\over 3\rho} \rho_\alpha\rho_\beta 
   \int_0^\infty r^3 dr\, 
  {\partial \beta V_{\alpha\beta}\over \partial r} g_{\alpha\beta}(r).
\label{R-virial}
\end{equation}
This expression cannot be applied to hard spheres, since the potential is 
discontinuous. In this case we have 
\begin{equation}
Z_{cc} = {16\pi R_c^3\over 3\rho} \rho^2_c g_{cc} (2 R_c).
\end{equation}
Eq.~(\ref{R-virial}) is also not convenient in the polymer-colloid case as 
$\beta V_{pc}(r)$ diverges as $r\to 0$. 
In the HNC case, this problem can be overcome by rewriting 
Eq.~(\ref{R-virial}) as 
\begin{equation}
Z_{\alpha\beta} = {2\pi \over 3\rho} \rho_\alpha\rho_\beta 
\int_0^\infty r^3 dr\, {\partial e^{-\beta V_{\alpha\beta}}\over \partial r}
   e^{h_{\alpha\beta}(r) - c_{\alpha\beta}(r)}.
\end{equation}
A similar formula can be analogously obtained in the 
case of the RY closure. 

Another quantity we shall be interested in is the isothermal compressibility 
$\kappa_T$ that can be either computed by using the virial route 
\begin{equation}
{\beta \over \kappa_T} = 
   \left( {\partial \beta P^{\rm (vir)} \over 
           \partial \rho_p}\right)_{\rho_c} \rho_p + 
   \left( {\partial \beta P^{\rm (vir)} \over 
           \partial \rho_c}\right)_{\rho_p} \rho_c
\label{chit-vir}
\end{equation}
or as \cite{BenNaim}
\begin{equation}
{\beta\over \kappa_T} = 
\rho - \sum_{\alpha\beta} \rho_\alpha \rho_\beta \hat{c}_{\alpha\beta}(0).
\label{chit-c}
\end{equation}
The two expressions are thermodynamically equivalent. However, when an 
approximate closure is used, two different results are obtained, as 
a consequence of the thermodynamic inconsistency of the approach. 
In the RY case, the parameter $\chi$ is fixed so that the two different
routes provide the same result for $\kappa_T$.

Finally, we shall consider the structure factors 
\begin{equation}
S_{\alpha\beta}(k) = \delta_{\alpha\beta} + 
    \sqrt{\rho_\alpha\rho_\beta} \hat{h}_{\alpha\beta}(k),
\end{equation}
and the concentration structure factor
\begin{equation}
 S_c(k) = x_p x_c \left[x_p S_{cc}(k) + x_c S_{pp}(k) - 
    2 \sqrt{x_p x_c} S_{cp}(k)\right].
\end{equation}
For $k\to 0$, $1/S_c(k)\to \partial^2 \beta g(x_p,P)/\partial x_p^2$, where 
$g(x_p,P)$ is 
the Gibbs free energy per particle. Hence, its divergence signals the 
thermodynamic instability of the homogeneous phase.

\section{Results} \label{sec3}

In order to solve the coupled integral equations, 
the correlation functions
are discretized on a regular grid. We usually take a step size 
$\Delta r/R_g = 10^{-3}$ and truncate the correlation functions
at $R_{\rm max}/R_g = N \Delta r$, with $N = 32768$. 
As we discuss in appendix \ref{AppA}, these choices make truncation and 
discretization errors
negligible. We use the standard Picard iterative method, which converges 
quite fast, except close to the termination line. We improve convergence 
by considering a mixing parameter $\alpha$. If $c^{(n)}_{\rm ini}(r)$  and
$c^{(n)}_{\rm end}(r)$ indicate
the direct correlation functions at the beginning and at 
the end of the $n$-th step of the iterative procedure, respectively, we set 
$c^{(n+1)}_{\rm ini}(r) = (1 - \alpha) c^{(n)}_{\rm ini}(r) + \alpha
c^{(n)}_{\rm end}(r)$. Far from the termination line, $\alpha$ is not a
relevant parameter. However, close to the termination line,
 convergence is only obtained if $\alpha$ is small. In some cases,
we took $\alpha \sim 10^{-2}$.

\subsection{Termination lines} \label{sec3.1}

In order to determine the termination line,
we work as
follows. We fix the colloid volume fraction $\Phi_c$ ($\Phi_c = 4 \pi R_c^3
N_c/3 V$, 
where $N_c$ is the number of colloids present in the box of volume $V$)
and solve the equations for a small value of the polymer density. 
Typically, if $\Phi_p = 4 \pi R_g^3 N_p/3 V$ ($N_p$ is the number of 
polymers present in the box of volume $V$), we start at 
$\Phi_p \approx 0.005$
for $\Phi_c\gtrsim 0.2$ and at $\Phi_p \approx 0.01$ for smaller
colloid volume fractions. Then, we increase $\Phi_p$ by steps 
$\Delta \Phi_p$. 

For $q = 1$ we have been able to increase $\Phi_p$ up to 2.5 for all values
of $\Phi_c \le 0.45$: We always find a regular solution of the 
integral equations.
This is not surprising,
as, for this value of $q$, Monte Carlo simulations indicate that the 
fluid-fluid binodal either does not exist or is located at quite large
values of the polymer volume fraction. In particular, Ref.~\cite{DMPP-15}
found no phase transition up to 
$\Phi_p = 2.12$, 1.73, 1.33 for $\Phi_c = 0.1,0.2,0.3$, respectively.

\begin{figure}[t]
\begin{center}
\begin{tabular}{cc}
\epsfig{file=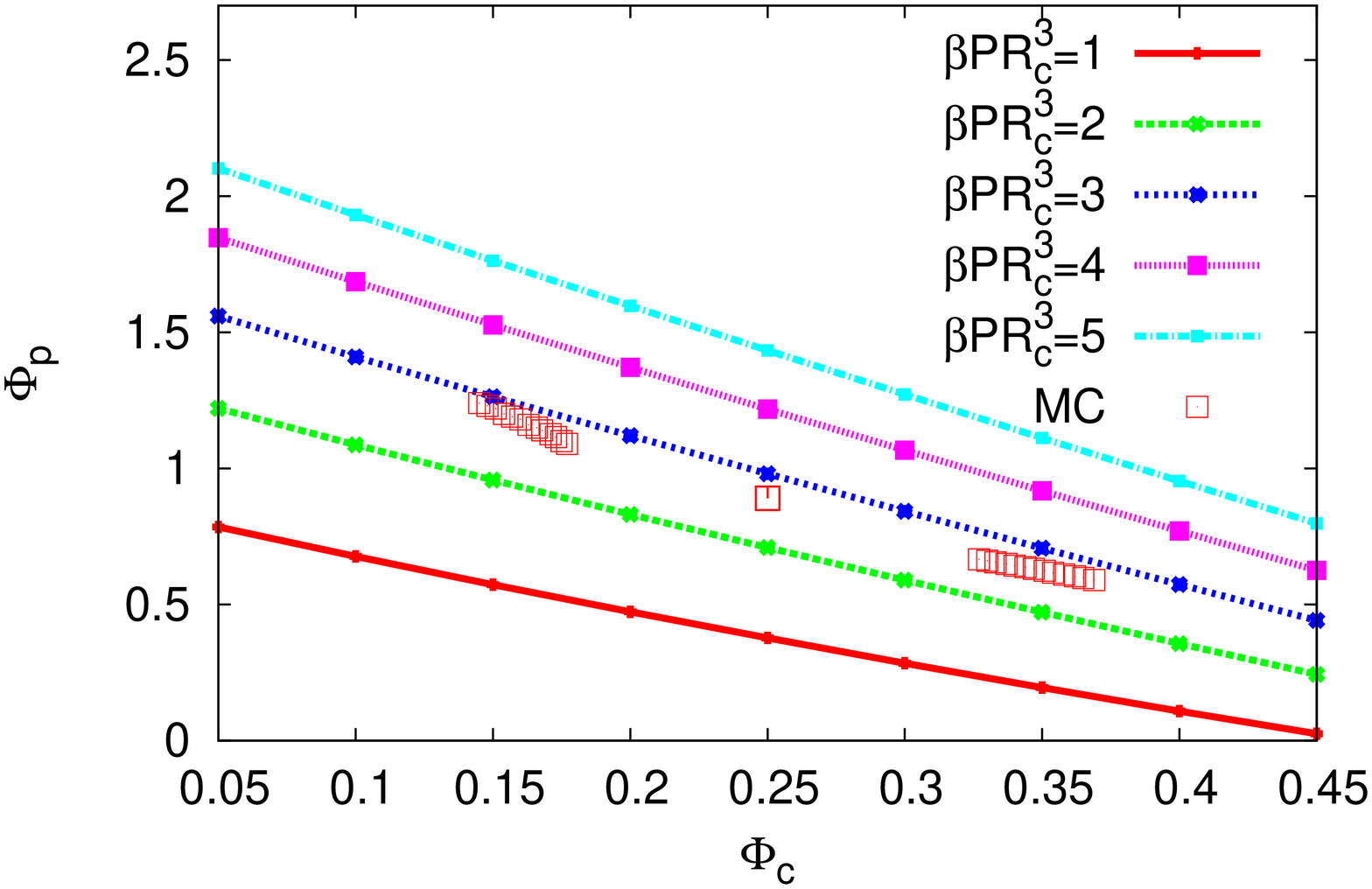,width=4.8truecm,angle=-90} 
   \hspace{0truecm} &
\epsfig{file=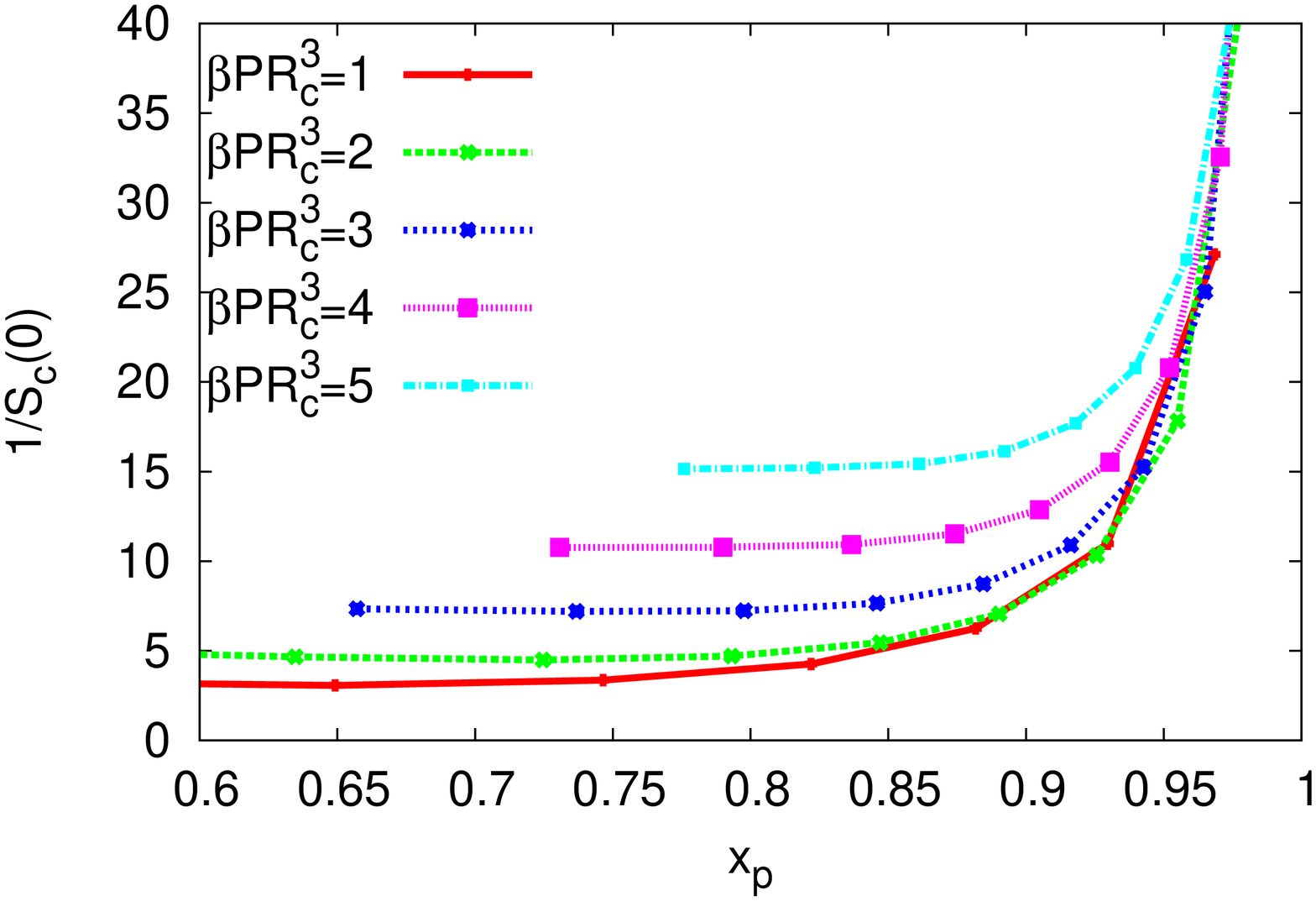,width=4.8truecm,angle=-90} 
\hspace{0truecm} \\
\end{tabular}
\end{center}
\caption{Results for $q = 0.8$ and the HNC/PY closure (in this case there is 
no termination line). Left: isobars corresponding to 
$\beta P R_c^3 = 1$, 2, 3, 4, and 5
in the $\Phi_c$, $\Phi_p$ plane. We also report the Monte Carlo estimates
of the binodal (empty squares, MC). Note that isobars 
with $\beta P R_c^3 \ge 3$ go through the two-phase region.
Right: $1/S_c(k=0)$ along the same isobars
reported in the left panel as a function of $x_p = N_p/(N_p + N_c)$. 
}
\label{Scq0p8-HNCPY}
\end{figure}

For $q = 0.8$ integral equations do not show any singular behavior 
if the HNC/PY closure is used. Also in this case 
we have been able to solve the equations for any 
$\Phi_p \le 2.5$ and $\Phi_c \le 0.45$. No phase demixing is 
observed, as it is evident from the behavior of 
$1/S_c(k=0)$ along five different isobars shown in Fig.~\ref{Scq0p8-HNCPY}.
At the critical point $1/S_c(k=0)$ should vanish. Instead, it 
increases as the pressure $P$ is increased, with no indication 
of a zero for some values of $P$ and $x_p$. Apparently, the HNC/PY closure 
fails even in reproducing the qualitative behavior of the system.

\begin{table}[h]
\caption{For each $q$ and $\Phi_c$ (first two columns), 
we report the polymer volume fraction $\Phi_p$ at which integral
equations no longer converge to a physical solution 
for three different closures:
HNC, HNC/PY, and RY. In the last column we report the polymer volume fraction
$\Phi_p^{\rm bin}$ at which the binodal, as computed by Monte Carlo
simulations \cite{DMPP-15}, occurs. We have also computed the termination line
for the RHNC closure, for $q = 0.5$ and $\Phi_c = 0.3$: $\Phi_p = 0.104$.
}
\label{table:noconv0p5}
\begin{center}
\begin{tabular}{cccccc}
\hline\hline
$q$ & $\Phi_c$ & HNC &   HNC/PY &  RY & $\Phi_p^{\rm bin}$ \\
\hline
0.5 
& 0.10 &  0.87 &  $\ge  2.5$ &  0.88 &  0.69 \\
& 0.20 &  0.23 &  $\ge  2.5$ &  0.34 &  0.53 \\
& 0.30 &  0.090&  0.15       &  0.18 &  0.38 \\
& 0.40 &  0.036&  0.07       &  0.115&  0.255 \\ \hline
0.8 
& 0.10 &  $\ge  2.5$ &  $\ge  2.5$ &  $\ge  2$ &  \\
& 0.20 &  0.61       &  $\ge  2.5$ &  $\ge  2$ &  1.0 \\
& 0.30 &  0.175      &  $\ge  2.5$ &  0.39 &  0.75  \\
& 0.40 &  0.086      &  $\ge  2.5$ &  0.29&  $0.5\le \Phi_p \le 0.6$ \\
\hline\hline
\end{tabular}
\end{center}
\end{table}

\begin{figure}[t]
\begin{center}
\begin{tabular}{c}
\epsfig{file=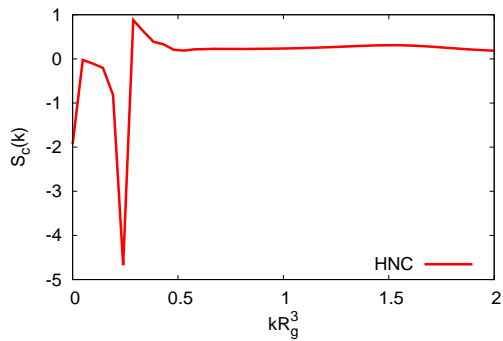,width=4.8truecm,angle=-90} 
   \hspace{0truecm} \\
\end{tabular}
\end{center}
\caption{Estimate of $S_c(k)$ for $\Phi_c = 0.3$ and $\Phi_p = 0.0905$,
on the termination line. Here $q = 0.5$ and we use the HNC closure.
}
\label{fig:Sck-termination}
\end{figure}

\begin{figure}[t]
\begin{center}
\begin{tabular}{c}
\epsfig{file=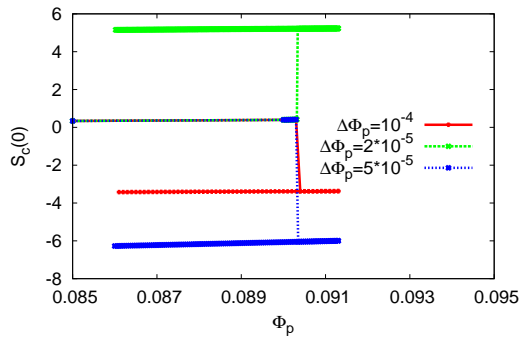,width=4.8truecm,angle=-90} 
   \hspace{0truecm} \\
\end{tabular}
\end{center}
\caption{Estimate of $S_c(k=0)$ for three different $\Delta \Phi_p$
and the HNC closure. 
We start from the values of $c_{\alpha\beta}(r)$ and $h_{\alpha\beta}(r)$
for $\Phi_p = 0.9$ and increase $\Phi_p$ by steps $\Delta \Phi_p$ 
up to $\Phi_p = 0.0915$, then we decrease $\Phi_p$ with the same schedule 
up to $\Phi_p = 0.086$. The termination line occurs for $\Phi_p = 0.0903$.
}
\label{fig:nonconv}
\end{figure}

For $q = 0.8$ a termination line is observed if we use the 
HNC or the RY closures, while for $q = 0.5$ a no-convergence domain is observed 
also by using the HNC/PY closure. Results for the 
termination lines for both values of $q$ are reported in 
Table~\ref{table:noconv0p5}. 
The termination line is determined as follows. Starting from the 
initial value $\Phi_p^{(0)}$, we subsequently solve the equations
for $\Phi_p^{(n)} = \Phi_p^{(0)} + n \Delta \Phi_p$, starting 
the iterations for the $n$-th density from the solution at $\Phi_p^{(n-1)}$. 
If $\Delta \Phi_p$
is large or the mixing parameter in the Picard iterations is of order 1,
we end up at a density $\Phi_p^{(M)}$ where the iterations no longer 
converge. Then, we consider again the solution at $\Phi_p^{(M-1)}$,
but now we significantly decrease $\Delta \Phi_p$ and the mixing parameter
(typically we take a parameter as small as 0.01). If we increase
again $\Phi_p$, we now observe that the Picard iterations always converge. 
However, at a very specific value of $\Phi_p$ the stable solution is no
longer physical, as $S_c(k)$ becomes
discontinuous at a finite value of $k$.
We identify the termination line as the smallest polymer density at
which $S_c(k)$ (the same occurs for all structure factors $S_{\alpha\beta}(k)$)
develops a discontinuity. An example is shown 
in Fig.~\ref{fig:Sck-termination}, where  we 
report $S_c(k)$ for $q=0.5$, $\Phi_c = 0.3$, $\Phi_p = 0.0903$, as obtained 
by using the HNC closure.  It is interesting to observe that while 
the position of the termination line is independent of the protocol 
used to increment $\Phi_p$, the singular solution depends on 
$\Delta \Phi_p$. For instance, in Fig.~\ref{fig:nonconv} we show the
estimates of $S_c(k=0)$ as a function of $\Phi_p$ for three different 
values of $\Delta \Phi_p$. Incrementing $\Phi_p$, 
at the termination line $\Phi_p = 0.0903$ we always observe a jump
in $S_c(0)$ to a new value. However, such value depends on $\Delta \Phi_p$. 
If we further
increase $\Phi_p$ beyond the termination line and then decrease again $\Phi_p$,
the unphysical solution appears to be stable: The structure factor 
changes smoothly with $\Phi_p$. Moreover, once $\Phi_p$ is again below
the termination-line value, if we use a small mixing parameter, we always 
obtain the unphysical solution.

\begin{figure}[t]
\begin{center}
\begin{tabular}{cc}
\epsfig{file=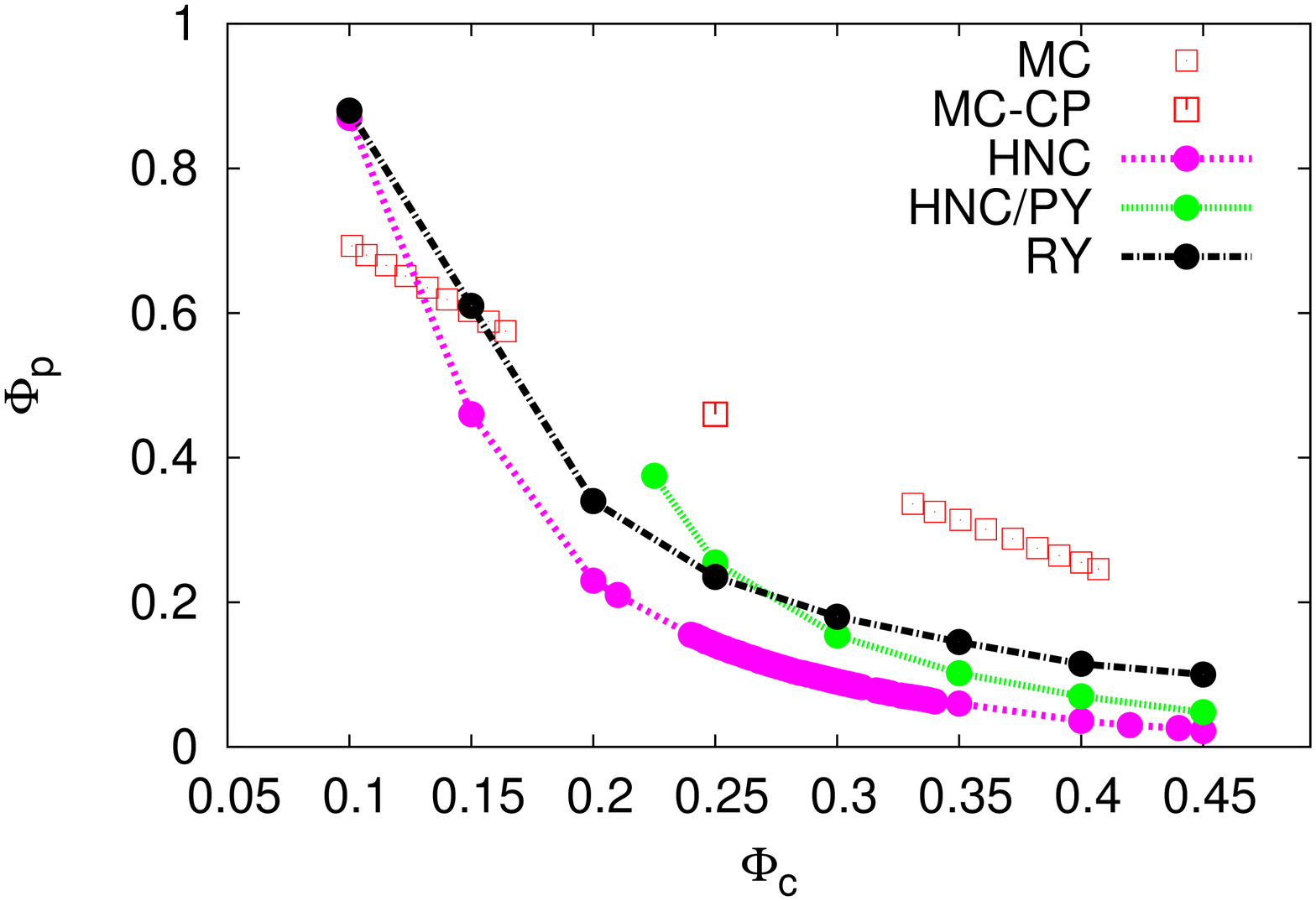,width=5truecm,angle=-90} 
   \hspace{0truecm} &
\epsfig{file=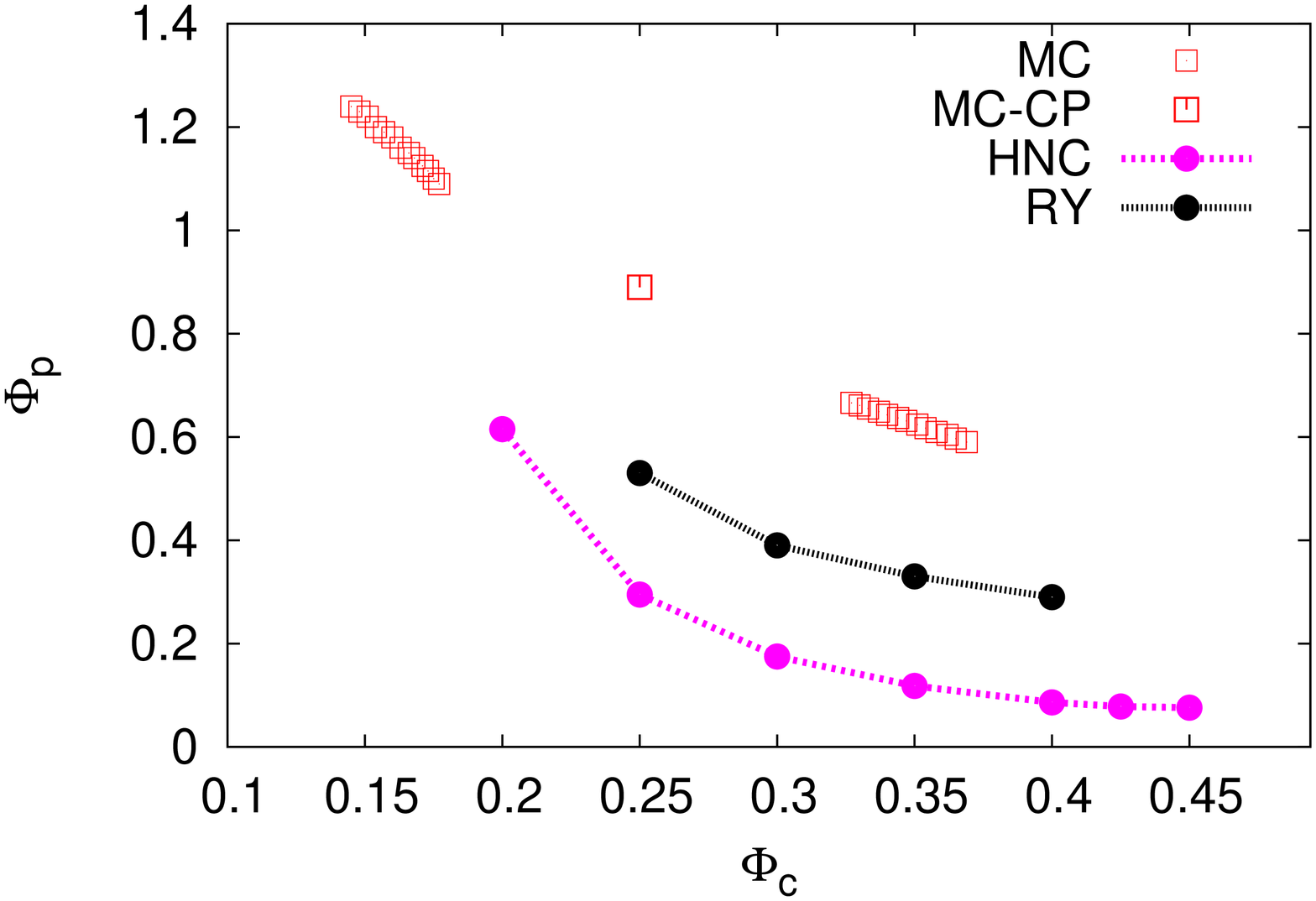,width=5truecm,angle=-90} 
\hspace{0truecm} \\
\end{tabular}
\end{center}
\caption{Phase diagram for $q = 0.5$ (left) and $q = 0.8$ (right).
We report the binodals obtained by Monte Carlo simulations (MC), 
the corresponding critical point (MC-CP), and 
the termination lines for each of the closures.
}
\label{Phase_diag}
\end{figure}

The termination lines for the different closures are reported in 
Fig.~\ref{Phase_diag}. In general, we find that the RY closure 
performs better than the HNC one, which stops converging at very small
values of $\Phi_p$ in the colloid-liquid phase. In all cases, however,
the termination line is significantly below the correct binodal, 
especially in the colloid-liquid phase $\Phi_c \gtrsim 0.25$. Clearly,
the convergence to an unphysical solution is not directly related to 
singularities in the thermodynamic behavior of the model. Therefore, 
the termination line provides a very poor approximation of the phase-separation 
line.

Let us finally consider the RHNC closure. Since this approach is quite 
complex, we have only analyzed one case: $q = 0.5$ and $\Phi_c = 0.3$. 
For $\Phi_p \approx 0$, the effective radius $R_p$ is equal to 
$0.837 R_g$. This is a completely reasonable value, indicating that 
polymers are effectively equivalent to hard spheres of radius approximately
equal to $R_g$. As $\Phi_p$ increases,
the effective radius $R_p$ decreases quite rapidly: for $\Phi_p = 0.1$ 
we find $R_p = 0.60 R_g$. Again, this is consistent with intuition, 
as we expect the polymer to shrink as $\Phi_p$ increases. Unfortunately,
we are not able to go much beyond $\Phi_p = 0.1$, as the RHNC equations
cease to converge at $\Phi_p = 0.104$. Hence, this approach represents
only a modest improvement with respect to the HNC approach
(the HNC termination line occurs at $\Phi_p = 0.090$).

\begin{figure}[t]
\begin{center}
\begin{tabular}{cc}
\epsfig{file=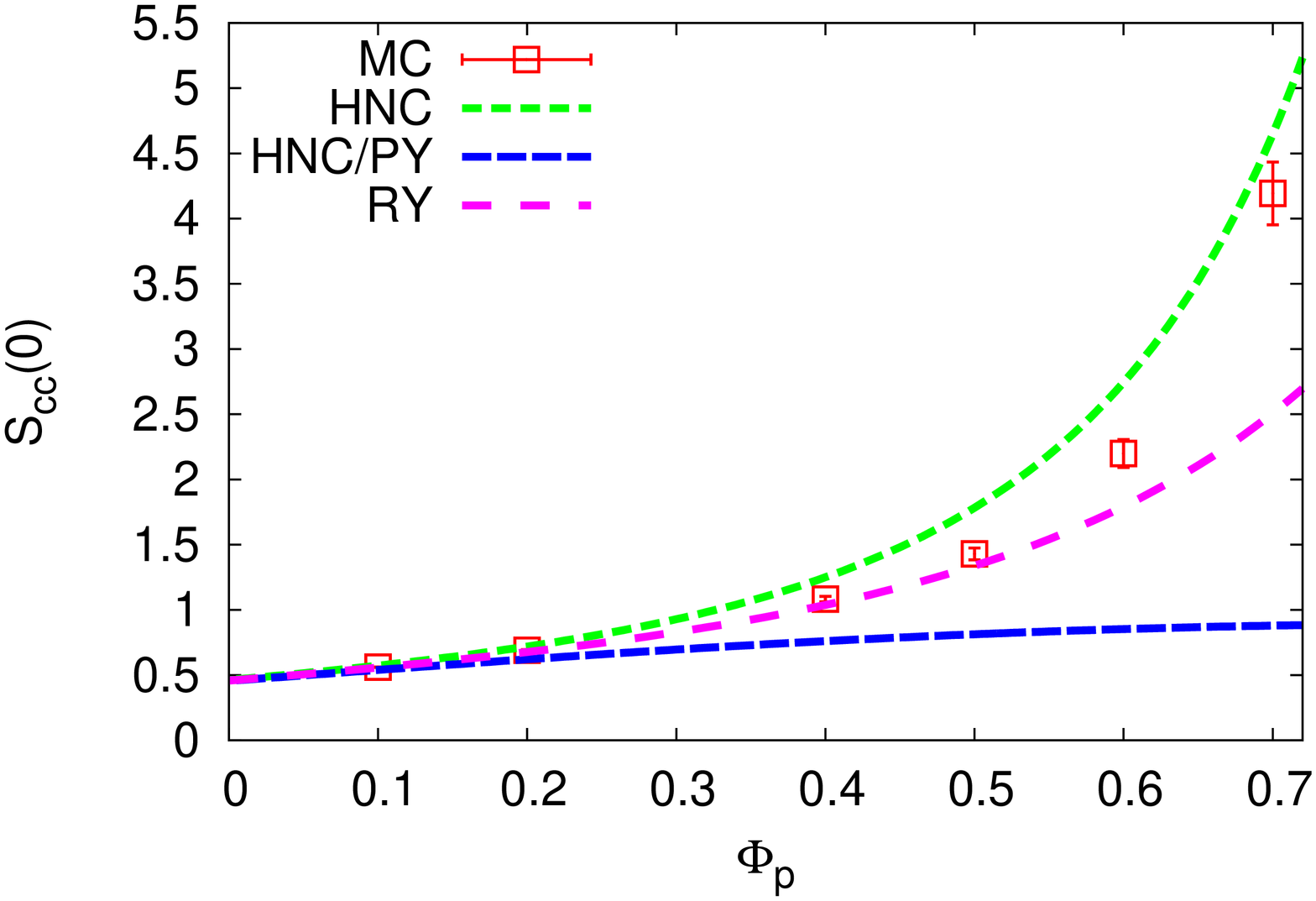,angle=0,width=5truecm,angle=-90} \hspace{0truecm} &
\epsfig{file=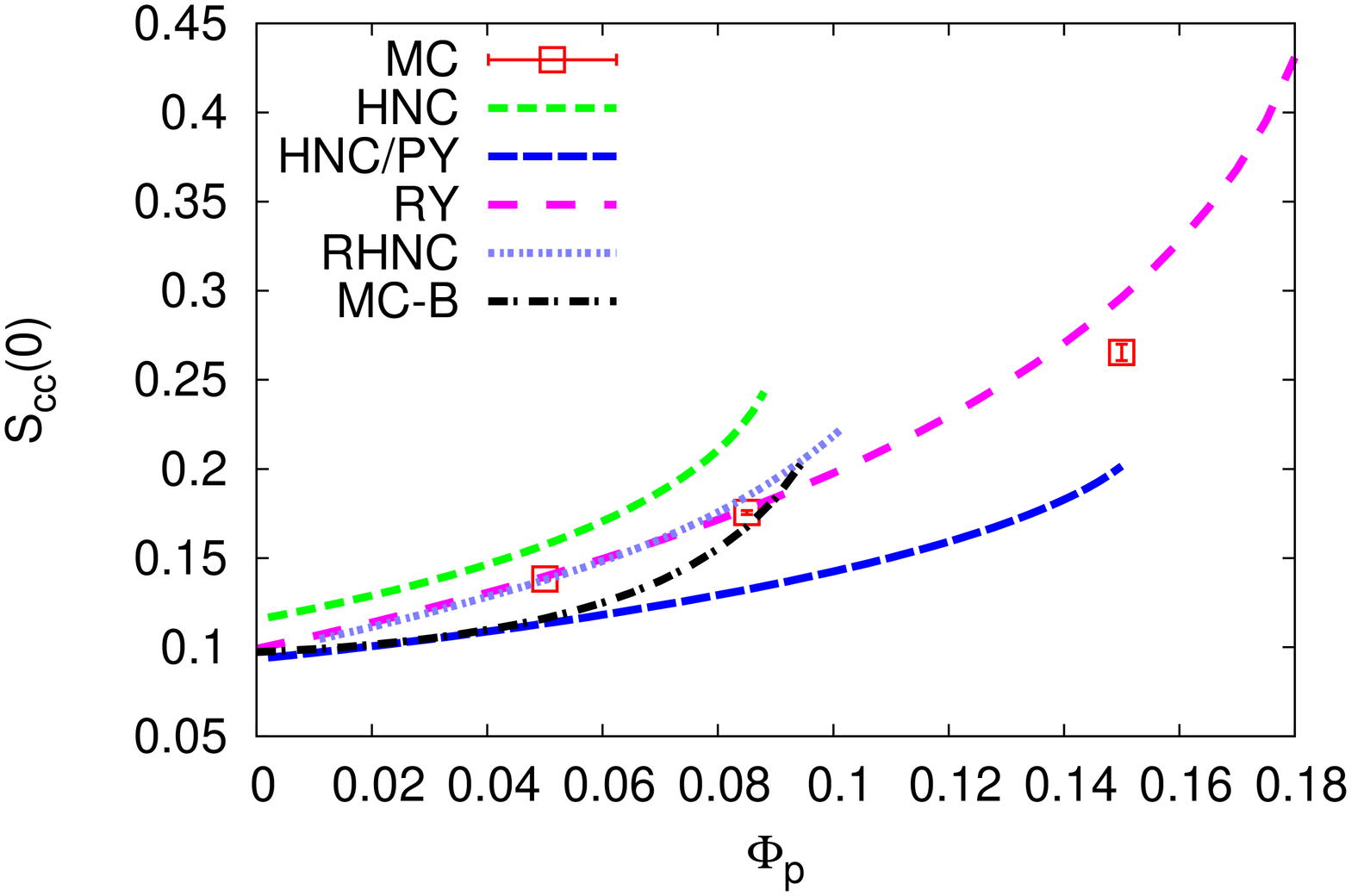,angle=0,width=5truecm,angle=-90} \hspace{0truecm} \\
\epsfig{file=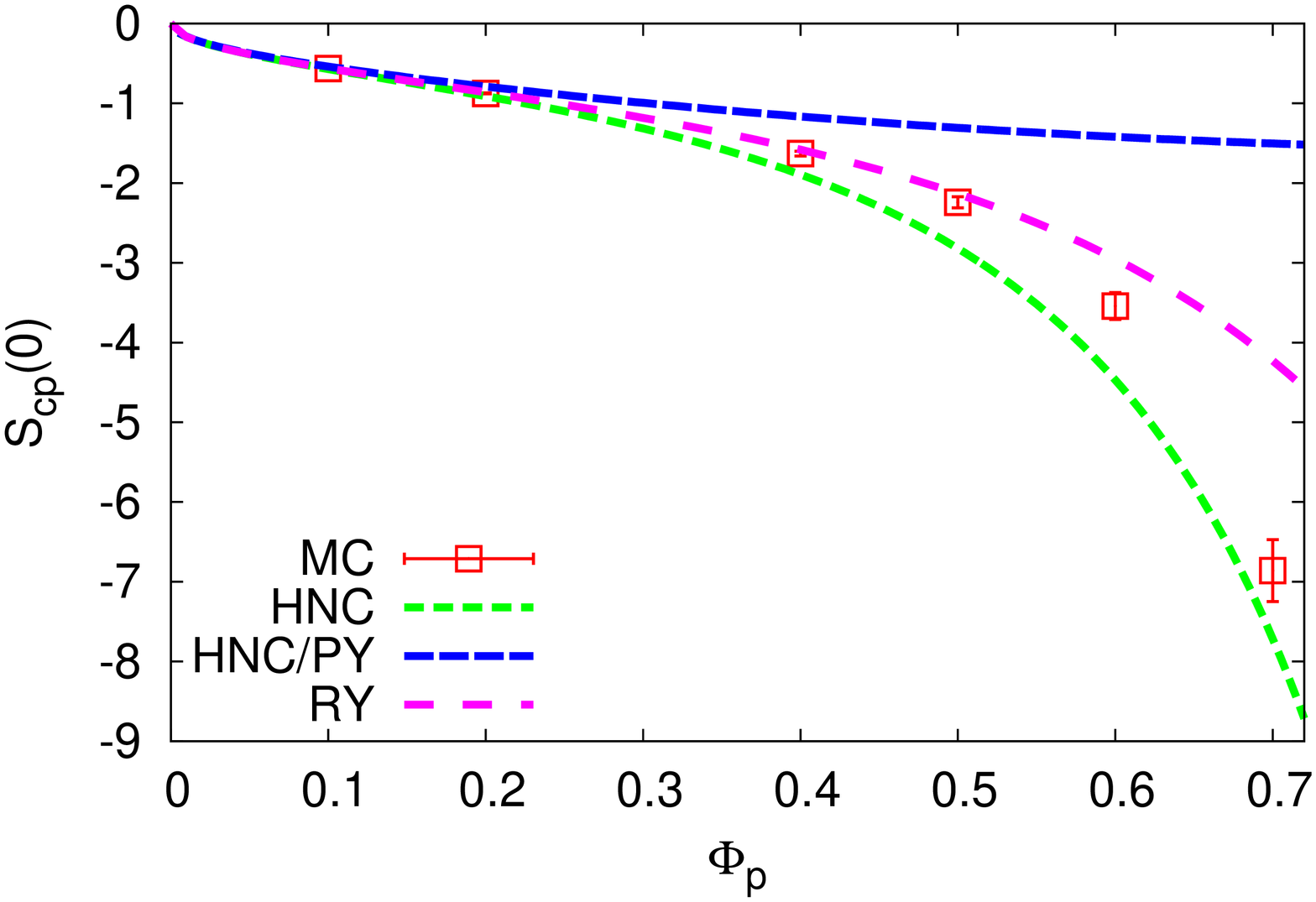,angle=0,width=5truecm,angle=-90} \hspace{0truecm} &
\epsfig{file=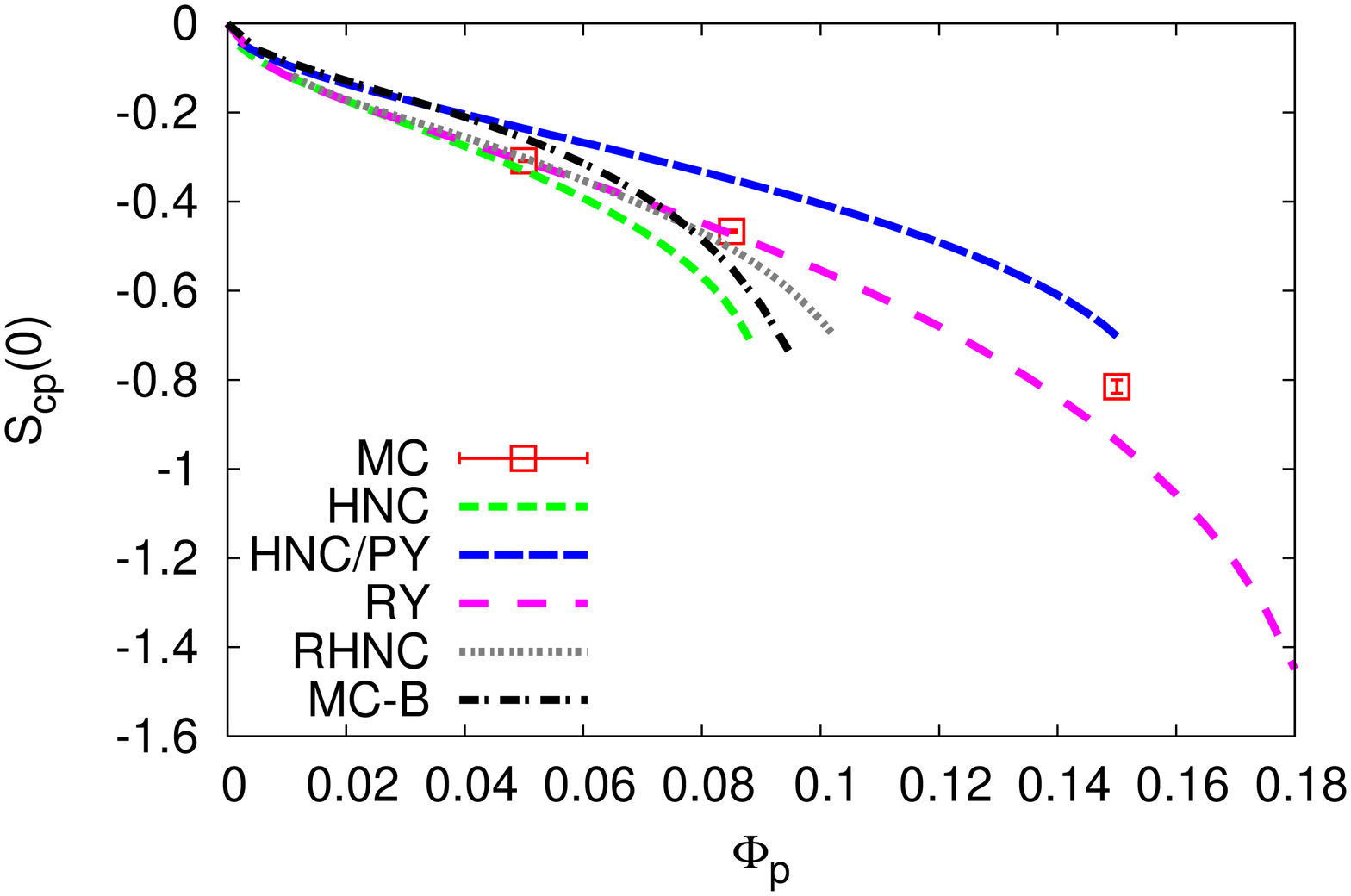,angle=0,width=5truecm,angle=-90} \hspace{0truecm} \\
\epsfig{file=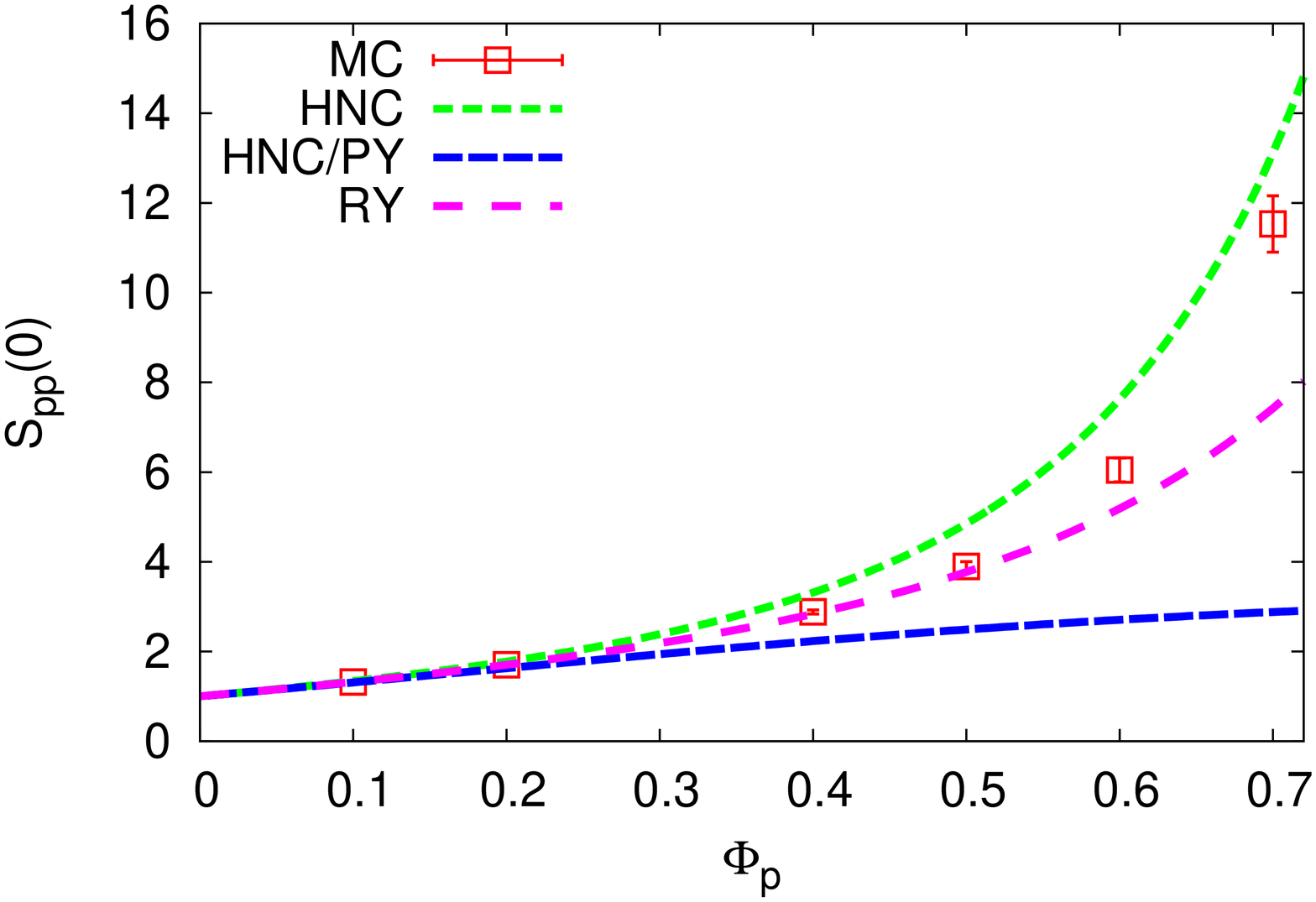,angle=0,width=5truecm,angle=-90} \hspace{0truecm} &
\epsfig{file=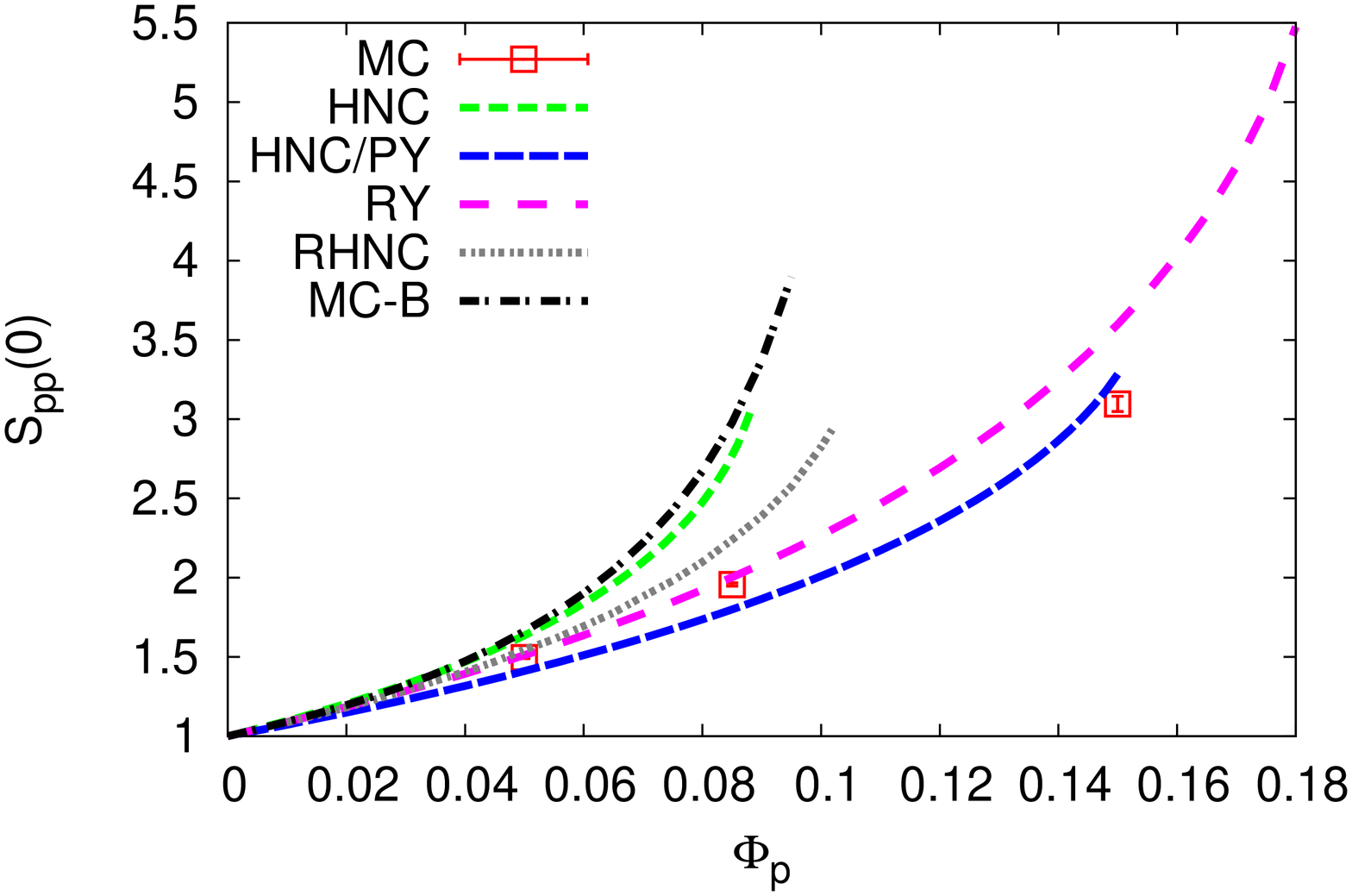,angle=0,width=5truecm,angle=-90} \hspace{0truecm} 
\end{tabular}
\end{center}
\caption{Structure factors $S_{\alpha\beta}(k=0)$ for $q= 0.5$ and 
$\Phi_c = 0.1$ (left) and $\Phi_c = 0.3$ (right). Lines are the 
results obtained by using the HNC, HNC/PY, and RY closures. Symbols are 
Monte Carlo data. For $\Phi_c = 0.3$ we also include results 
for the RHNC closure and results obtained by using the 
zero-polymer-density Monte Carlo bridge functions (MC-B), as 
discussed in Sec.~\ref{sec3.4}.
}
\label{Scq0.5}
\end{figure}

\subsection{Structural behavior in the homogeneous phase} \label{sec3.2}

We wish now to compare the integral-equation predictions with the Monte Carlo
ones in the homogeneous phase. We consider the case $q=0.5$, in which 
a termination line occurs for all considered closure relations. We 
begin by analyzing the 
structure factors $S_{\alpha\beta}(k=0)$, which are 
directly related to thermodynamics by the compressibility equations 
\cite{HansenMcDonald,BenNaim}. In Fig.~\ref{Scq0.5} we report the 
corresponding estimates for two values of $\Phi_c$, $\Phi_c = 0.1$ and 0.3, 
that lie on opposite sides with respect to the critical point located at
$\Phi_{c,\rm crit} = 0.25$, $\Phi_{p,\rm crit} =0.46$, as 
estimated by Monte Carlo simulations \cite{DMPP-15}.

For $\Phi_c = 0.1$ the HNC/PY closure significantly underestimates the 
structure factors. Clearly, $|S_{\alpha\beta}(0)|$ increases too slowly
as $\Phi_p$ increases, explaining why convergence is observed at least up
to $\Phi_p = 2.5$, see Table~\ref{table:noconv0p5}. The HNC and RY 
estimates increase faster. The latter are more accurate than the HNC 
ones for small densities, but they significantly underestimate 
$|S_{\alpha\beta}(0)|$ close to the binodal, which is located at 
$\Phi_p \approx 0.70$ \cite{DMPP-15}. The fact that the RY results are less 
accurate than the HNC ones near the binodal may be surprising, as the RY closure is 
a generalization of the HNC closure. It simply indicates that the 
requirement of thermodynamic consistency does not  necessarily lead to more 
accurate results. Note that both HNC and RY integral equations also 
converge for some values of $\Phi_p$ in the 
metastable region beyond the binodal, see Fig.~\ref{Phase_diag}. 
In this domain the 
structure factors $S_{\alpha\beta}(0)$ are 
quite large [on the binodal, Monte Carlo simulations give
$S_{pp}(0) = 11.5(6)$, $S_{cp}(0) = -6.7(4)$, $S_{cc}(0) = 4.2(2)$].
Therefore, even though we do not observe an exact divergence of
$S_{\alpha\beta}(k=0)$,
for this value of $\Phi_c$ we can take the termination line 
as a good estimate of the spinodal. 

For $\Phi_c = 0.3$ the behavior is quite different and the termination lines 
occur at values of $\Phi_p$ significantly smaller than that of the binodal. 
Moreover, integral equations stop converging when the structure 
factors $|S_{\alpha\beta}(0)|$ 
are relatively small, at least if compared with the values they assume 
on the binodal at $\Phi_c = 0.1$. For instance, the HNC and HNC/PY 
equations both cease to converge when $S_{pp}(0) \approx 3$, 
while $S_{pp}(0)\approx 5$ on the RY termination line. 
Comparing the integral-equation estimates 
with the Monte Carlo results, we see that the RY closure is here the most 
accurate, in agreement with previous studies \cite{DLL-02,PP-14}, although 
it fails to converge well before the binodal.
As for the RHNC, the estimates of $S_{cc}(0)$ and $S_{cp}(0)$ are consistent
with the RY ones and the Monte Carlo data up to $\Phi_p \approx 0.08$. 
On the other hand, the RHNC estimates of $S_{pp}(0)$ increase
too fast for $\Phi_p \gtrsim 0.04$, looking similar to the 
HNC estimates. Also in this case the termination line occurs for 
$S_{pp}(0) \approx 3$. 

\begin{figure}[t]
\begin{center}
\begin{tabular}{cc}
\epsfig{file=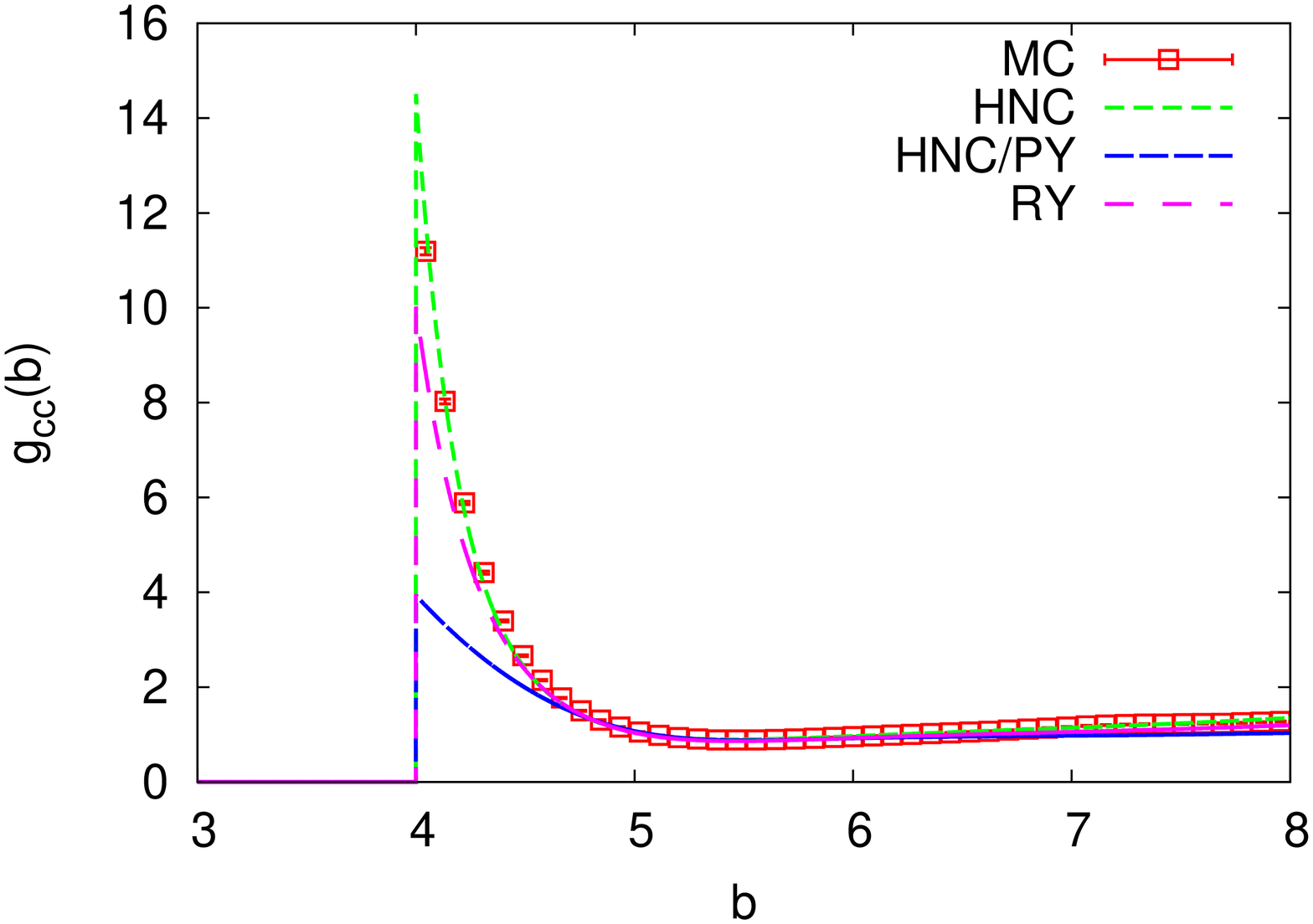,angle=0,width=5truecm,angle=-90} 
   \hspace{0truecm} &
\epsfig{file=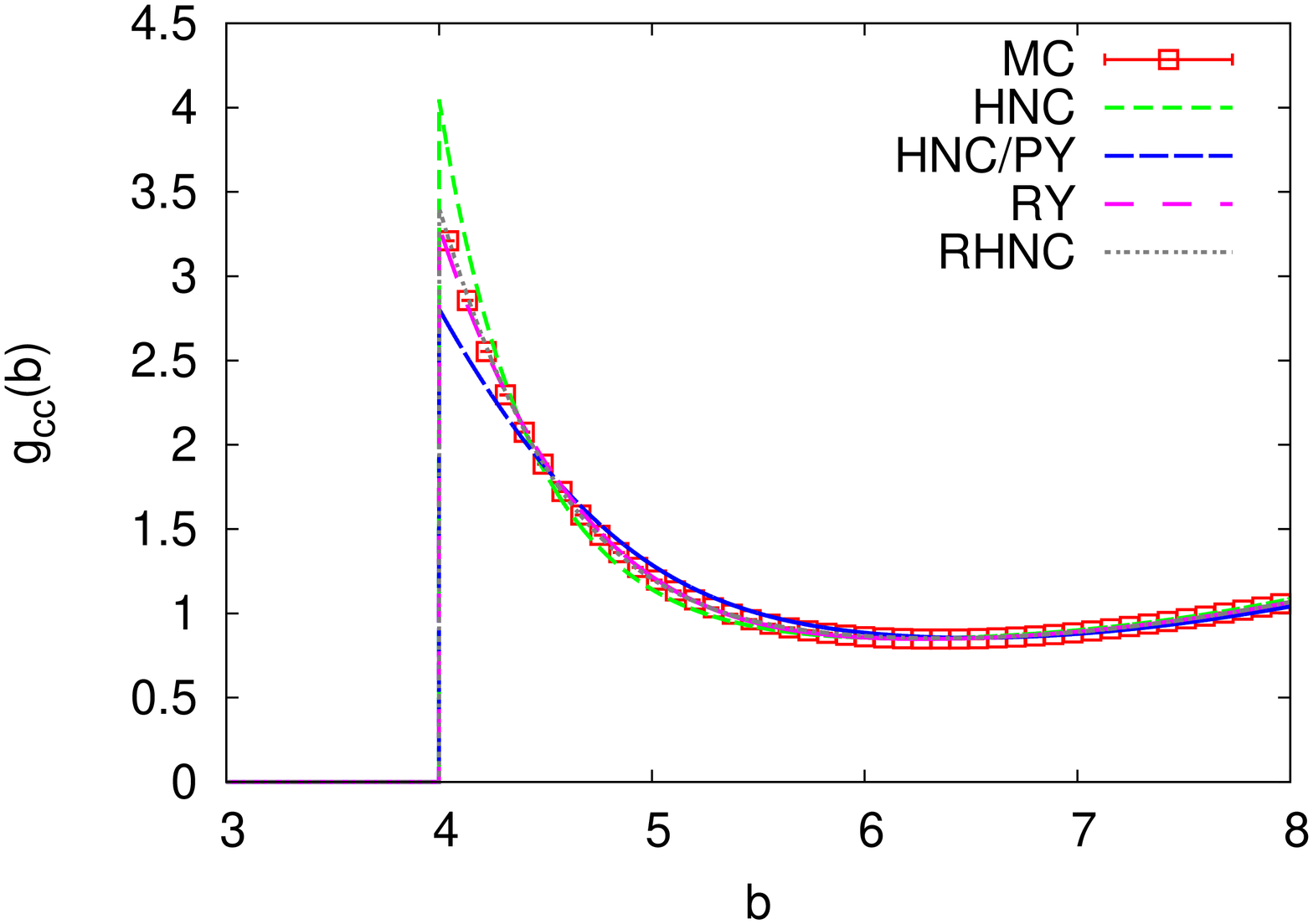,angle=0,width=5truecm,angle=-90} 
\hspace{0truecm} \\
\epsfig{file=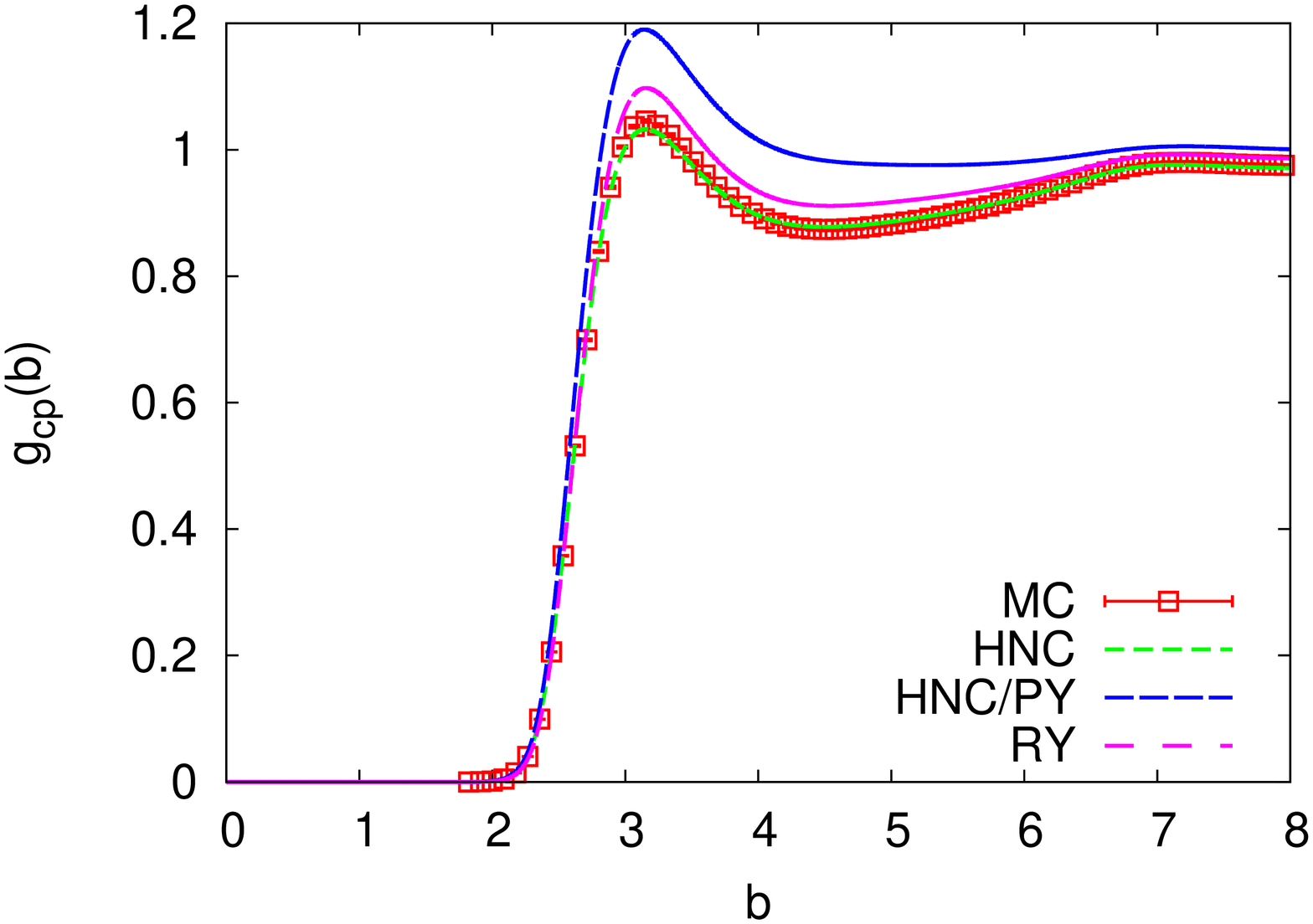,angle=0,width=5truecm,angle=-90} 
\hspace{0truecm} &
\epsfig{file=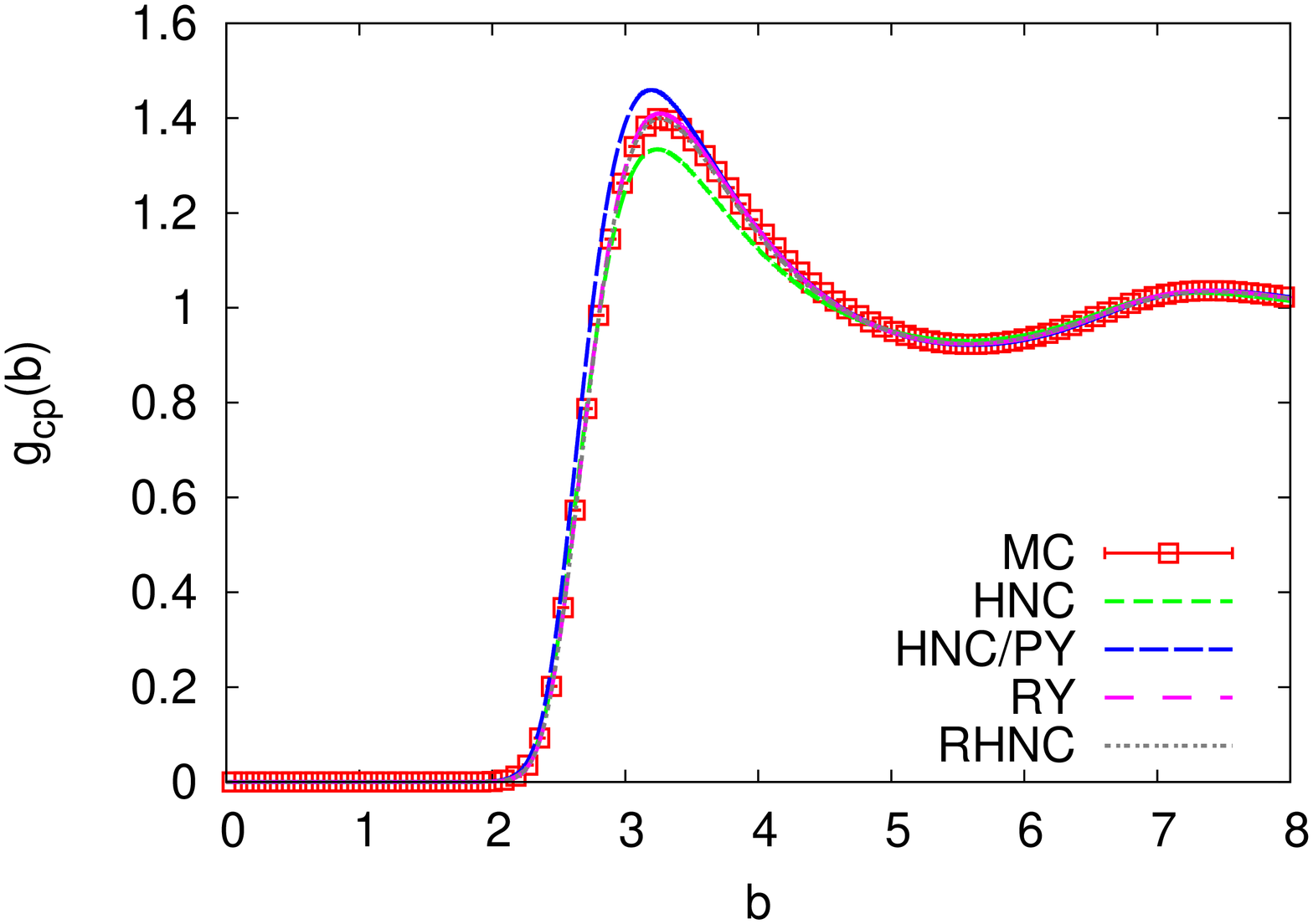,angle=0,width=5truecm,angle=-90} 
\hspace{0truecm} \\
\epsfig{file=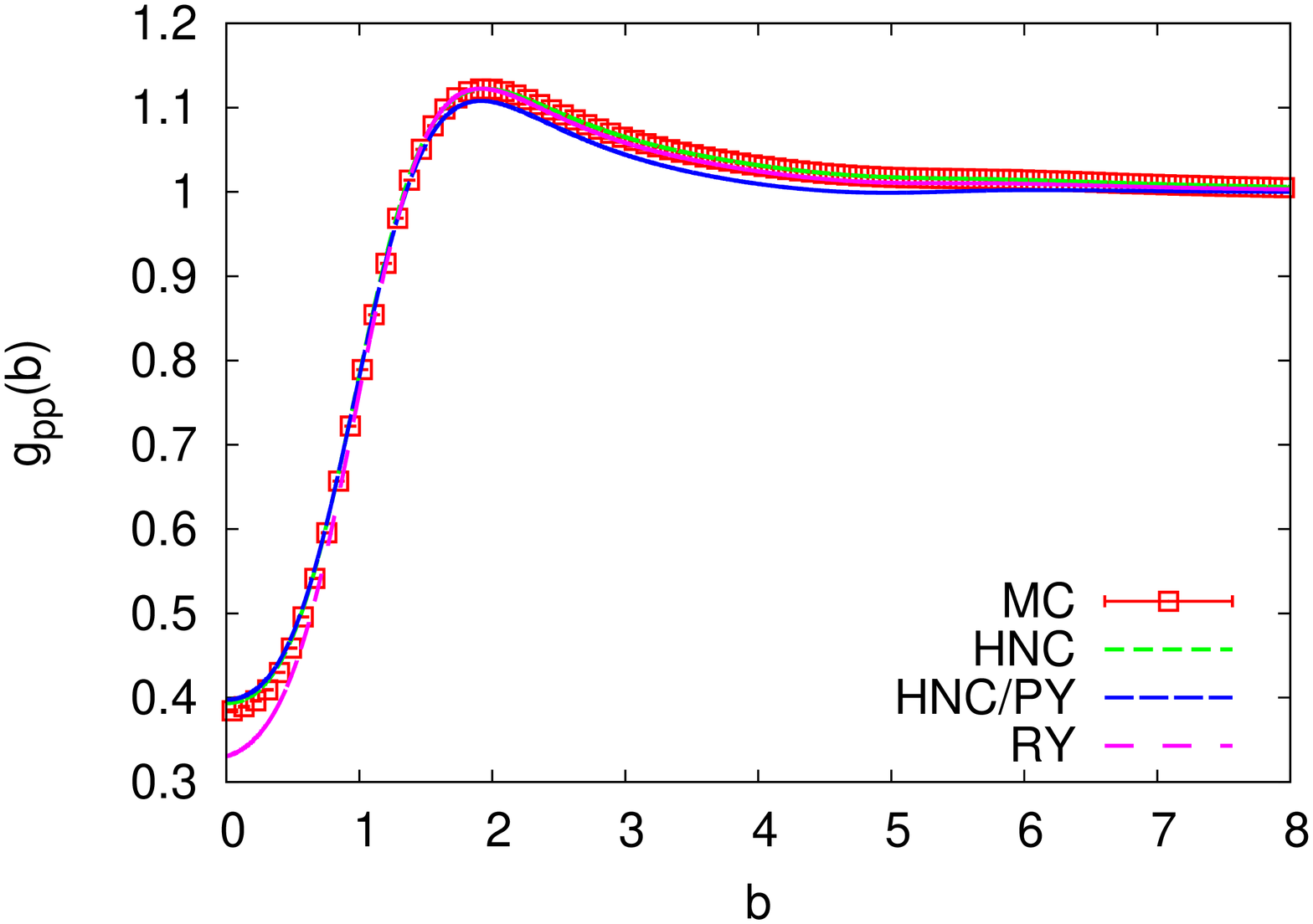,angle=0,width=5truecm,angle=-90} 
\hspace{0truecm} &
\epsfig{file=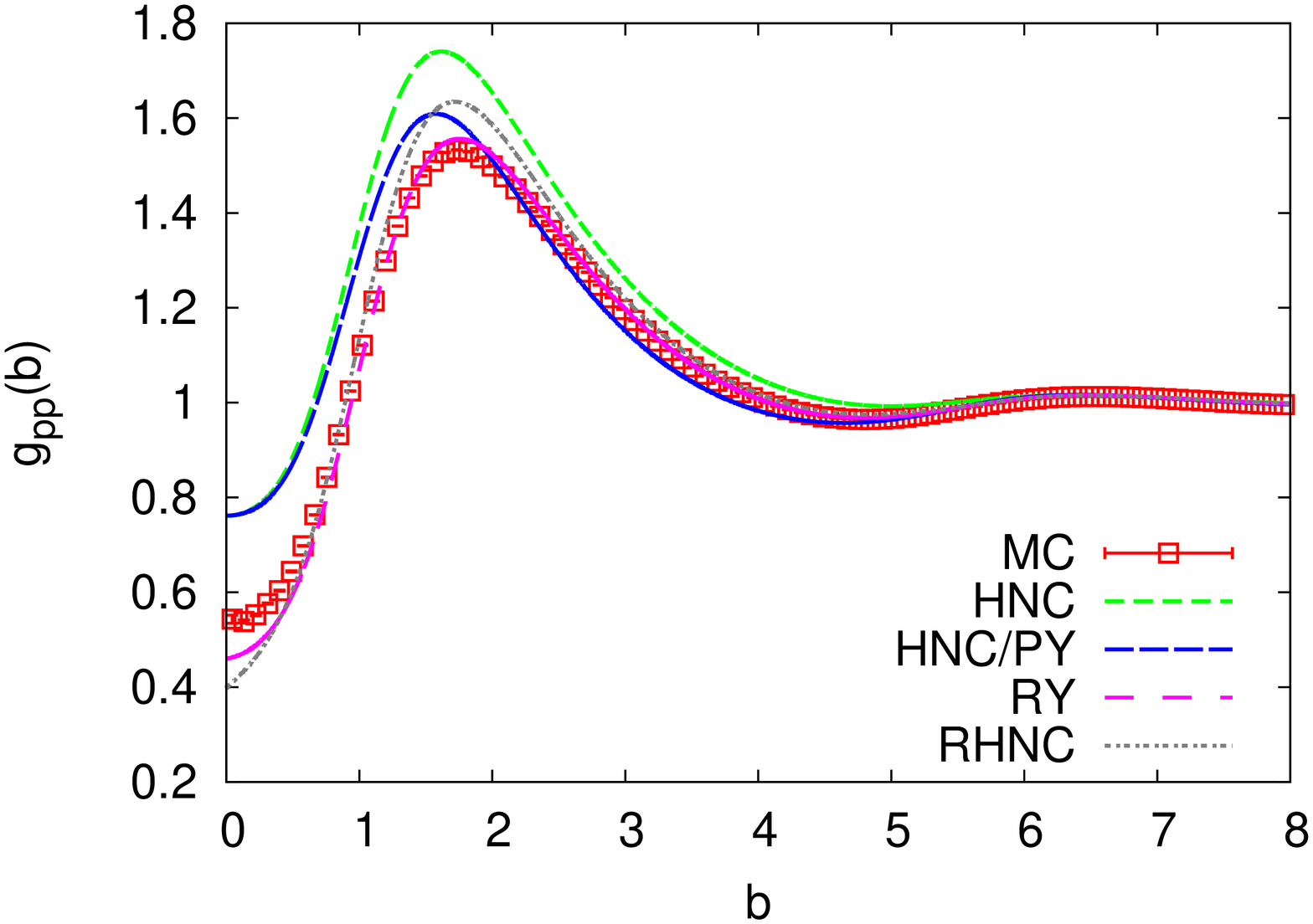,angle=0,width=5truecm,angle=-90} 
\hspace{0truecm} 
\end{tabular}
\end{center}
\caption{Pair correlation functions $g_{\alpha\beta}(r)$ as a function 
of $b = r/R_g$ for $q= 0.5$ at 
two different state points:
$\Phi_c = 0.1$, $\Phi_p = 0.70$ (left) and 
$\Phi_c = 0.3$, $\Phi_p = 0.085$ (right).
Lines are the results obtained by using the HNC, HNC/PY, RY, RHNC 
closures. Symbols are
Monte Carlo data. 
}
\label{gq0.5}
\end{figure}

\begin{figure}[t]
\begin{center}
\begin{tabular}{c}
\epsfig{file=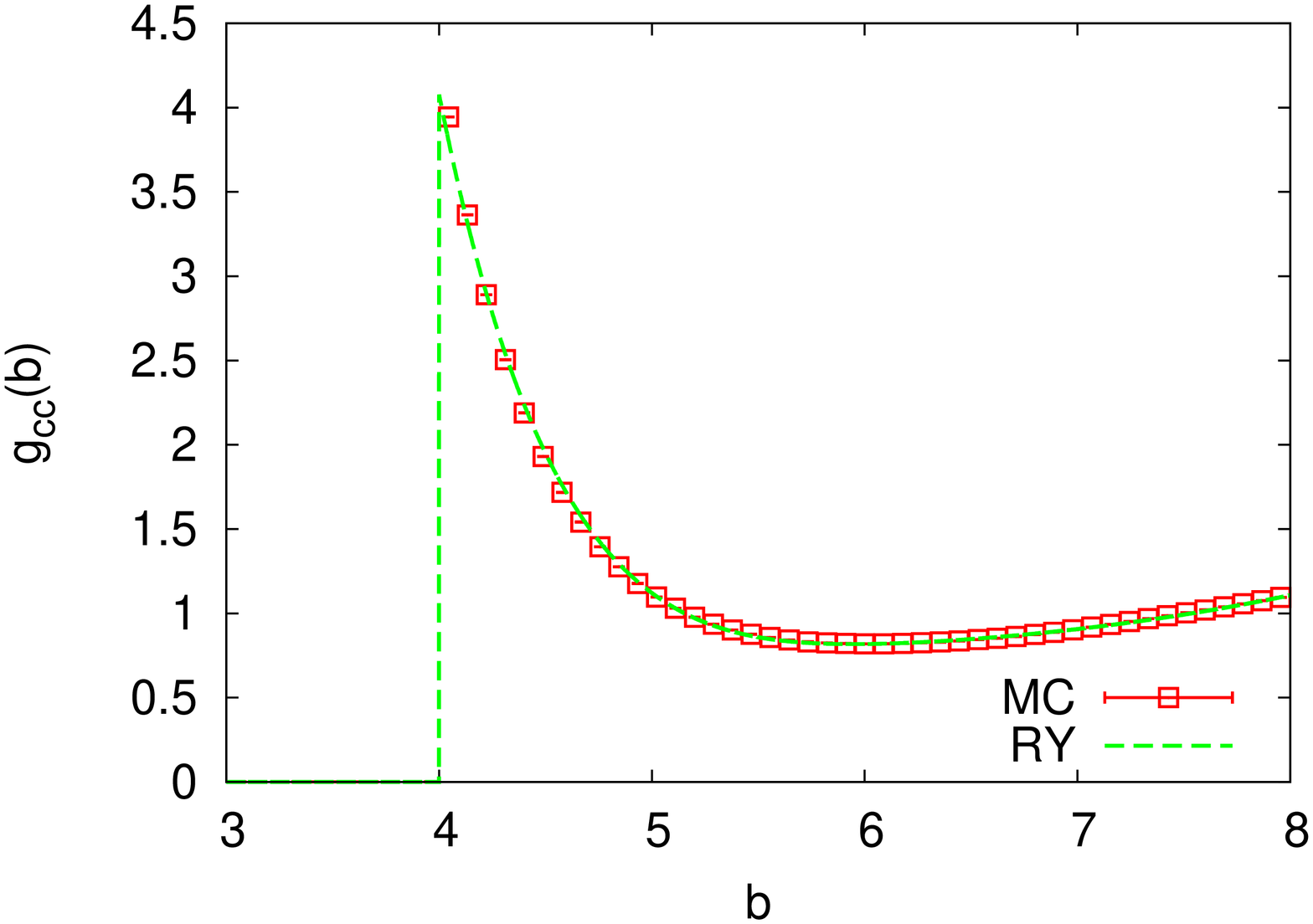,angle=0,width=5truecm,angle=-90} 
\hspace{0truecm} \\
\epsfig{file=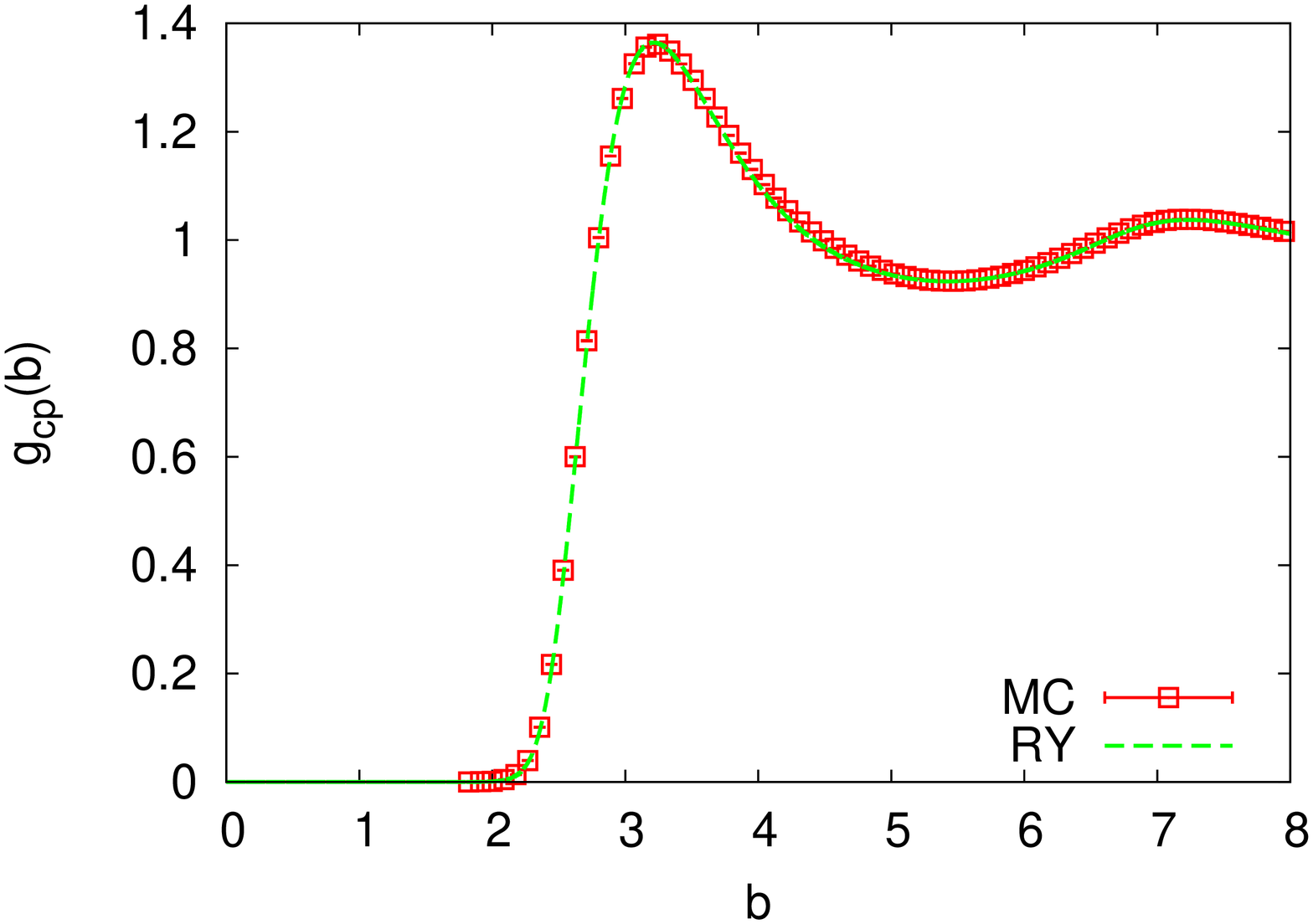,angle=0,width=5truecm,angle=-90} 
\hspace{0truecm} \\
\epsfig{file=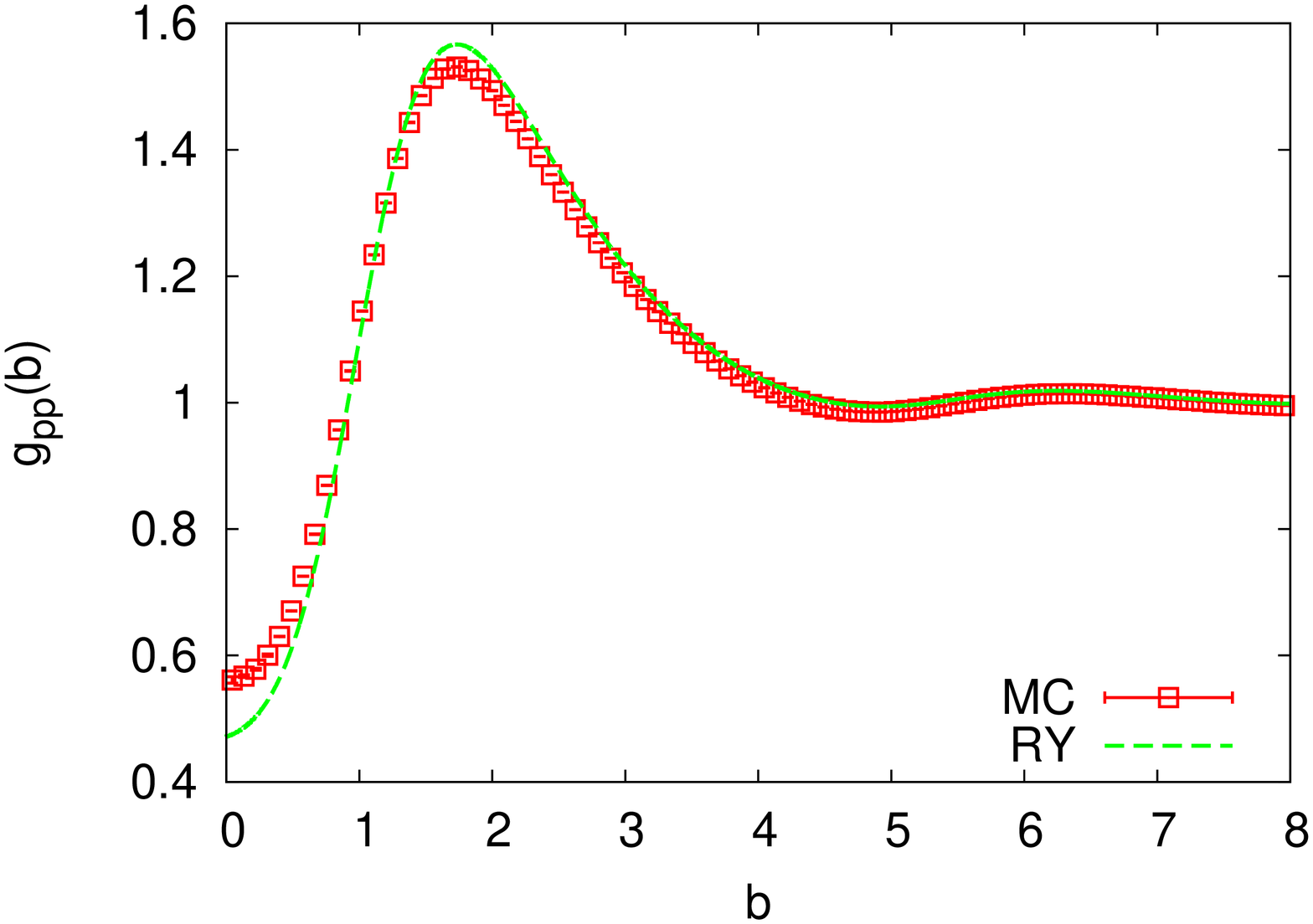,angle=0,width=5truecm,angle=-90} 
\hspace{0truecm} 
\end{tabular}
\end{center}
\caption{Pair correlation functions $g_{\alpha\beta}(r)$ as a function 
of $b = r/R_g$ for $q= 0.5$ at 
$\Phi_c = 0.3$, $\Phi_p = 0.15$.
Lines are the results obtained by using the RY
closure. Symbols are Monte Carlo data.  
}
\label{gq0.5-2}
\end{figure}

As a second test let us compare the pair distribution functions. 
For $\Phi_c = 0.1$ and $\Phi_p = 0.7$, i.e. on the binodal, 
see Fig.~\ref{gq0.5}, 
all closures reasonably reproduce the polymer-polymer distribution
function. Deviations are instead observed for the polymer-colloid
and especially for the colloid-colloid distribution function. 
The largest deviations are observed for the HNC/PY closure. For instance,
the colloid-colloid correlation is significantly 
underestimated at contact. While an extrapolation of the Monte Carlo
data predicts $g_{cc}(2 R_c) \approx 13$-14, we estimate 
$g_{cc}(2 R_c) \approx 4$ by using the HNC/PY closure. 
The RY closure performs better, although it is also unable to 
predict the correct value of $g_{cc}(r)$ at contact and slightly overestimates
$g_{cp}(r)$ at the first peak. As for the structure factors, the HNC closure is 
the most accurate one for this value of $\Phi_c$, as the HNC curves fall on
top of the Monte Carlo data.

At $\Phi_c = 0.3$ the behavior is quite different, see Fig.~\ref{gq0.5}. 
For $\Phi_p = 0.085$, 
close to the HNC termination line, HNC results are not accurate, especially
for $g_{pp}(r)$, which is significantly overestimated for $r \lesssim 2 R_g$. 
The value of $g_{cc}(r)$ at contact is also significantly overestimated. 
The HNC/PY closure gives results that are only marginally better than the HNC 
ones, while the RY estimates are in full agreement with the Monte 
Carlo data. The RHNC estimates of $g_{cc}(r)$ and $g_{cp}(r)$ are 
in agreement with the data, but this is not the case for $g_{pp}(r)$, which 
is overestimated for $1\lesssim r/R_g \lesssim 2$, the region in which 
the correlation function shows the first peak.
At $\Phi_p = 0.15$ we only have RY data, as integral equations no longer
converge for the other closures. The results are reported in
Fig.~\ref{gq0.5-2}. Pair correlations $g_{cc}(r)$ and $g_{cp}(r)$ 
are well reproduced, while relatively small deviations are observed 
for $g_{pp}(r)$. Apparently, RY estimates are relatively accurate 
even close to the corresponding termination line, located at 
$\Phi_p = 0.18$.

\subsection{Bridge functions at zero polymer density} \label{sec3.3}

\begin{figure}[t]
\begin{center}
\begin{tabular}{cc}
\epsfig{file=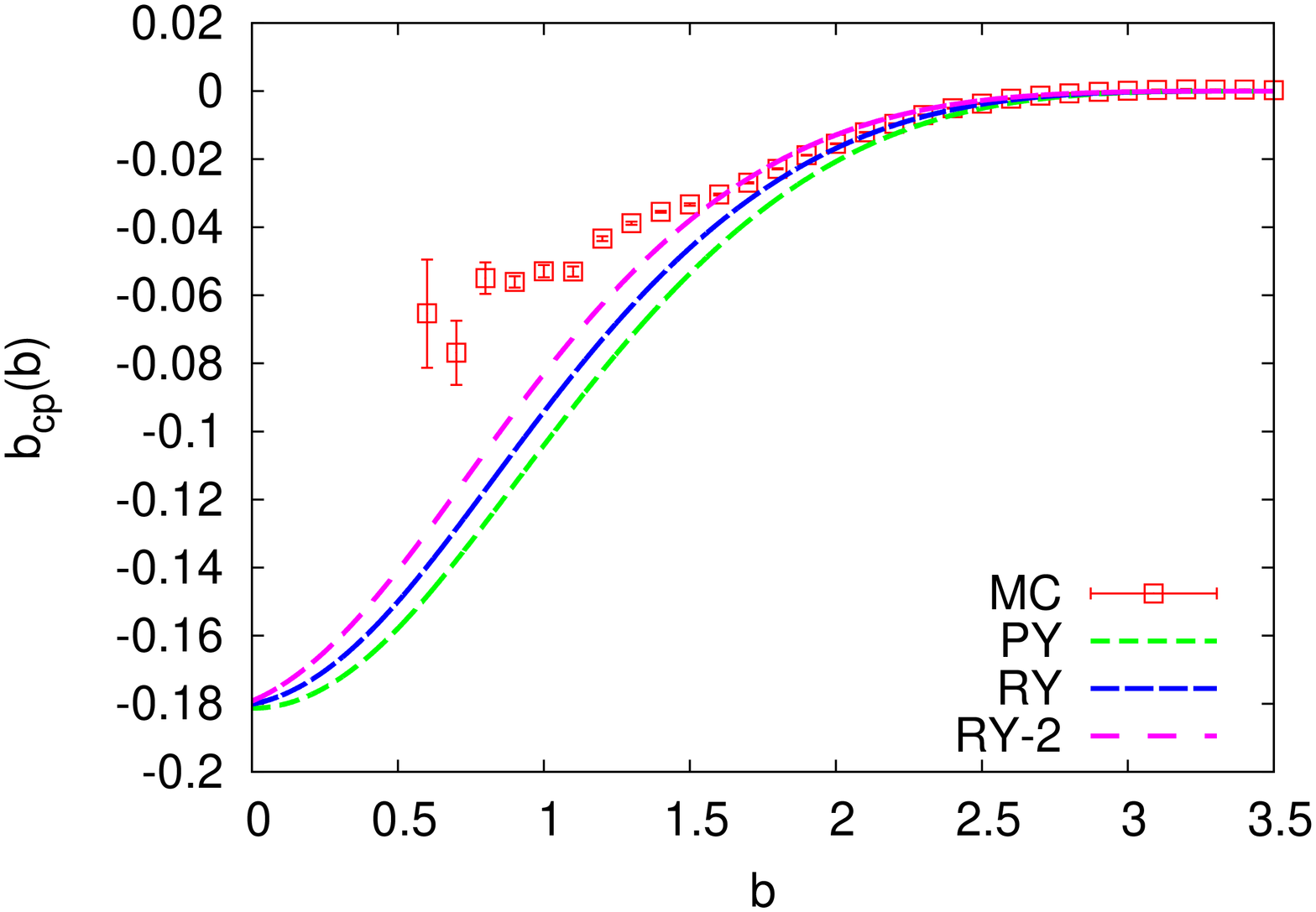,angle=0,width=4.8truecm,angle=-90} 
   \hspace{0truecm} &
\epsfig{file=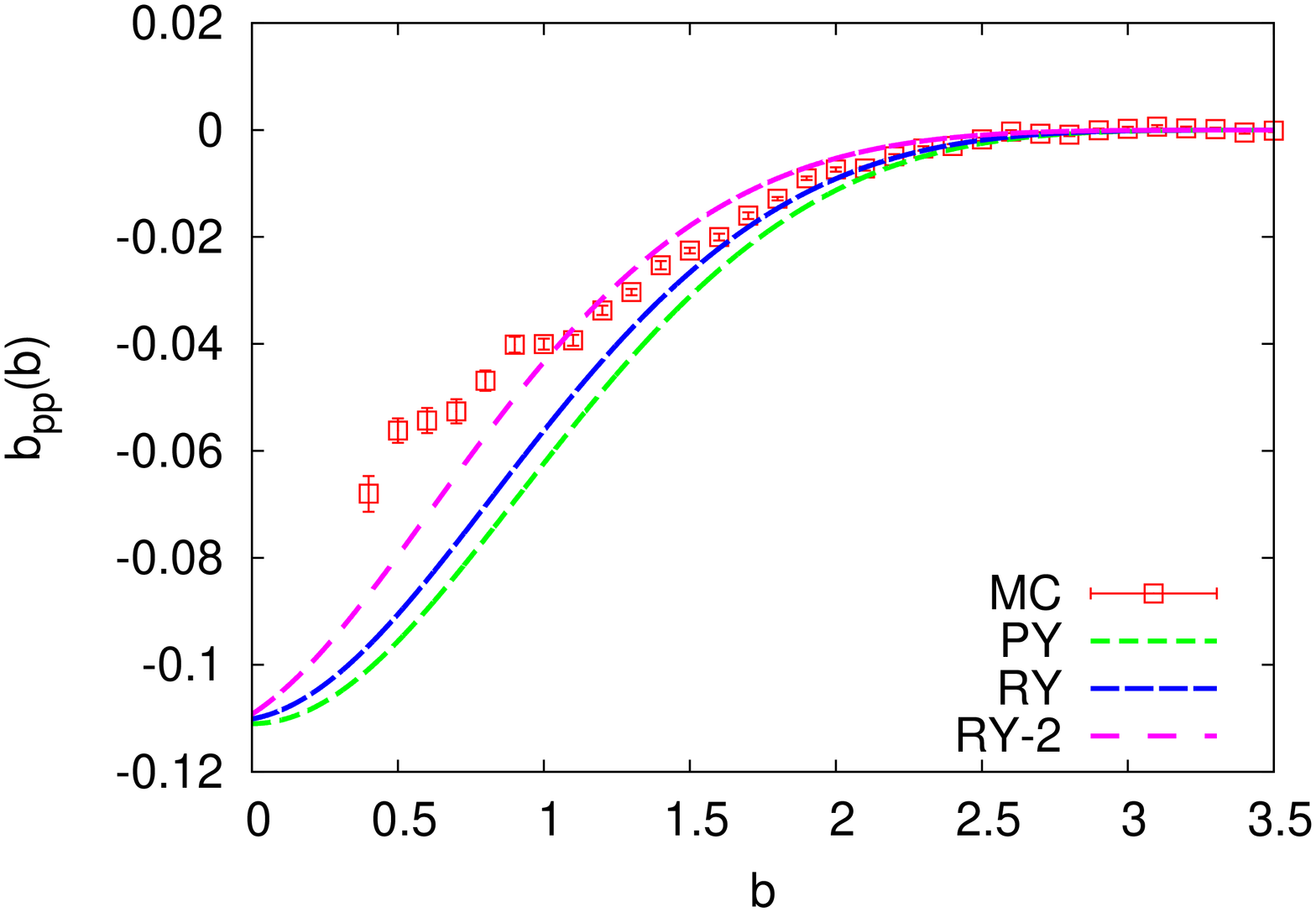,angle=0,width=4.8truecm,angle=-90}  \\
\epsfig{file=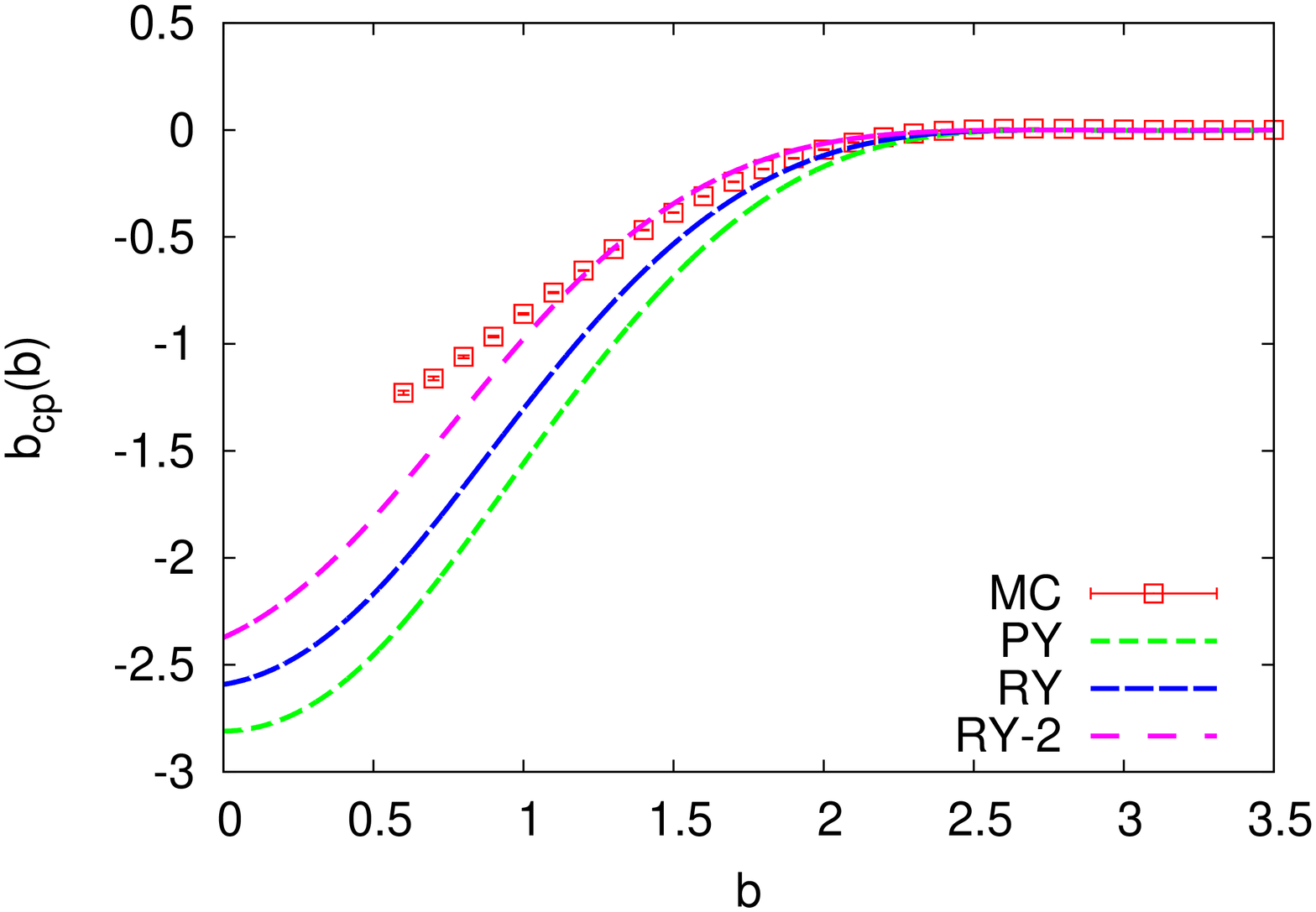,angle=0,width=4.8truecm,angle=-90} 
   \hspace{0truecm} &
\epsfig{file=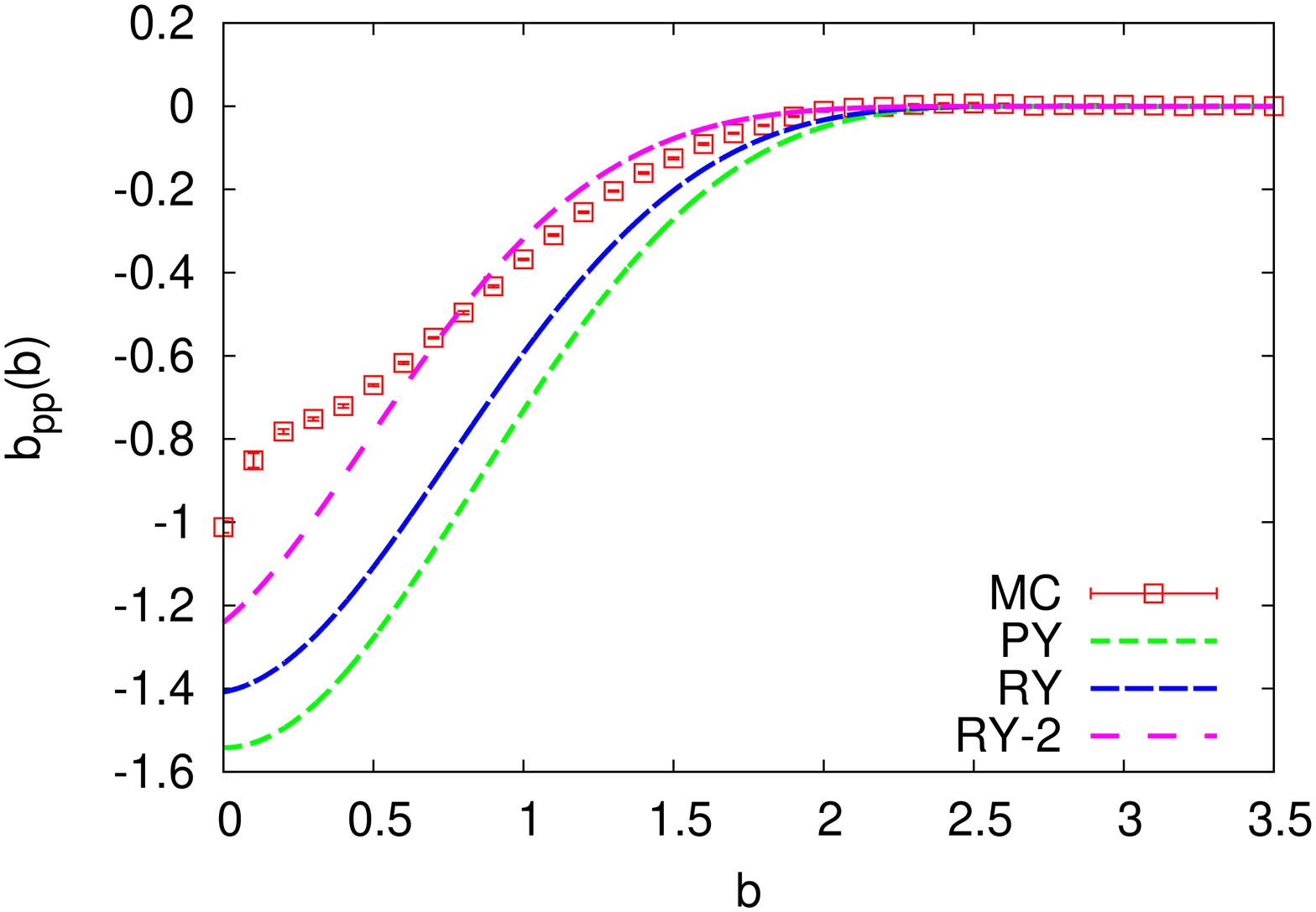,angle=0,width=4.8truecm,angle=-90}  \\
\end{tabular}
\end{center}
\caption{Bridge functions for $q=1$ as a function of $b = r/R_g$: 
on the left we report 
$b_{cp}(r)$, on the right $b_{pp}(r)$. Top: $\Phi_c = 0.1$;
bottom: $\Phi_c = 0.3$. 
We report the Monte Carlo estimates (MC) 
as well as those obtained by using the different closures. 
RY-2 labels the results obtained by using the two-parameter
RY closure discussed in the text.
}
\label{bridgeq1}
\end{figure}

\begin{figure}[t]
\begin{center}
\begin{tabular}{cc}
\epsfig{file=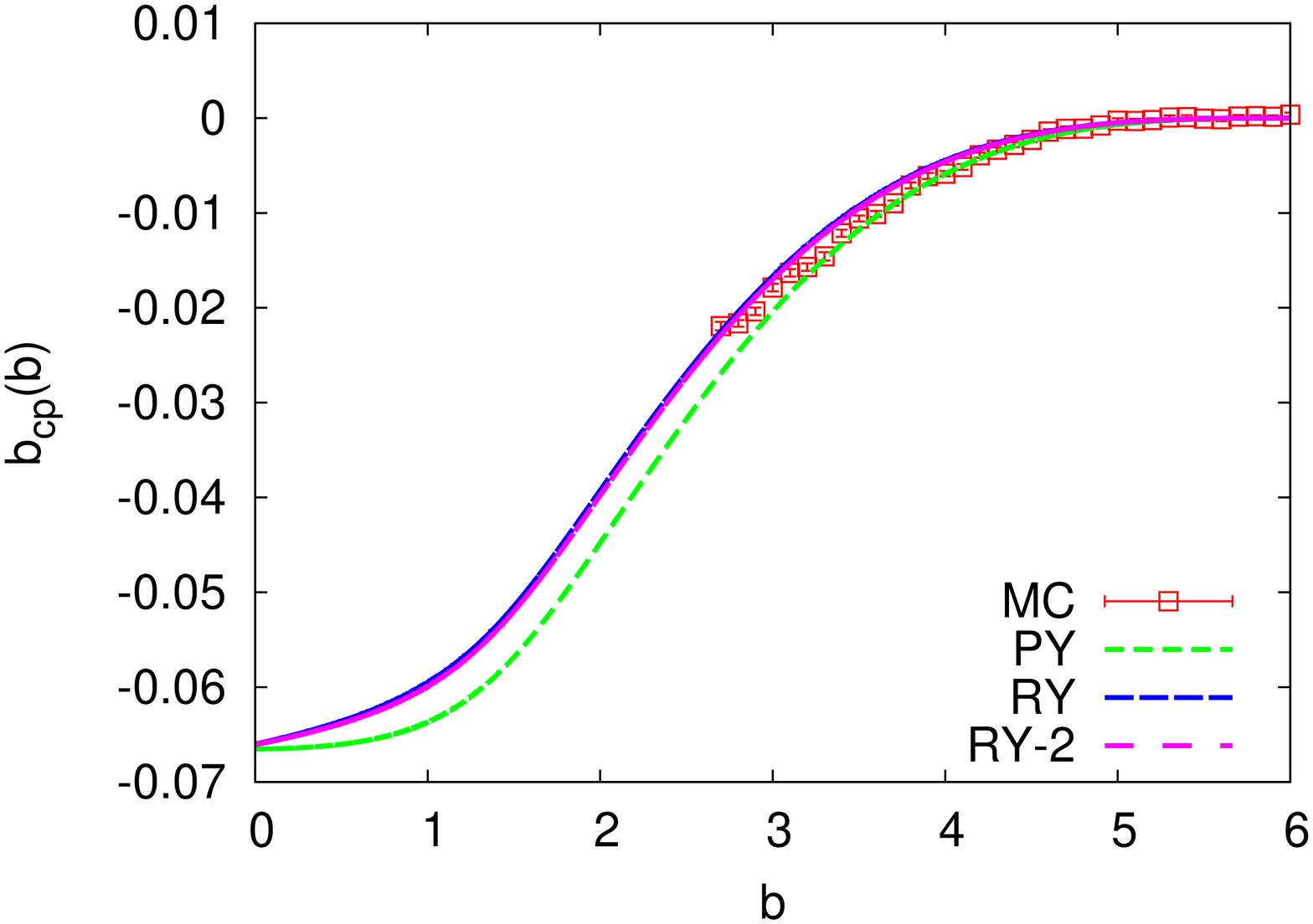,angle=0,width=4.8truecm,angle=-90} 
   \hspace{0truecm} &
\epsfig{file=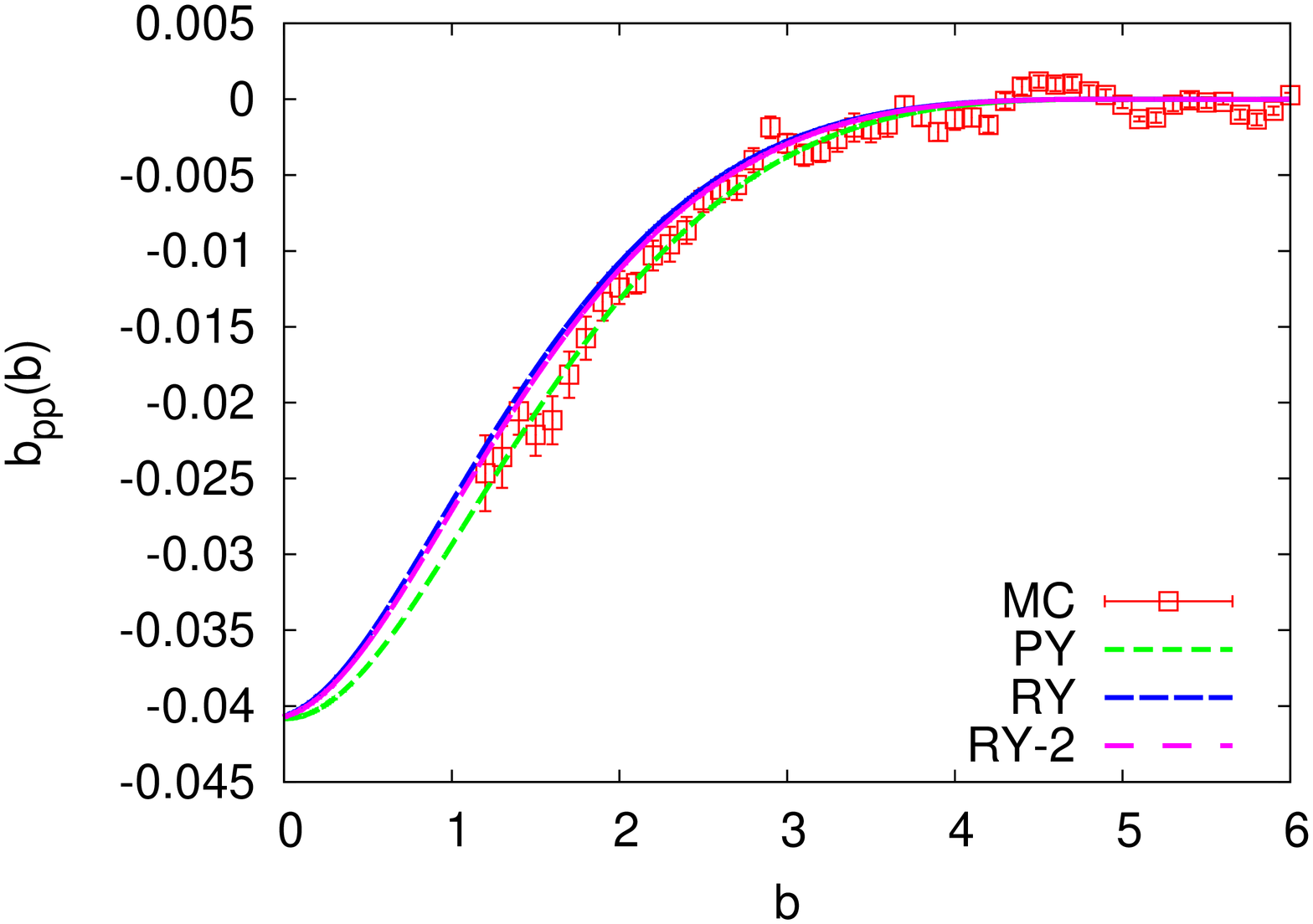,angle=0,width=4.8truecm,angle=-90}  \\
\epsfig{file=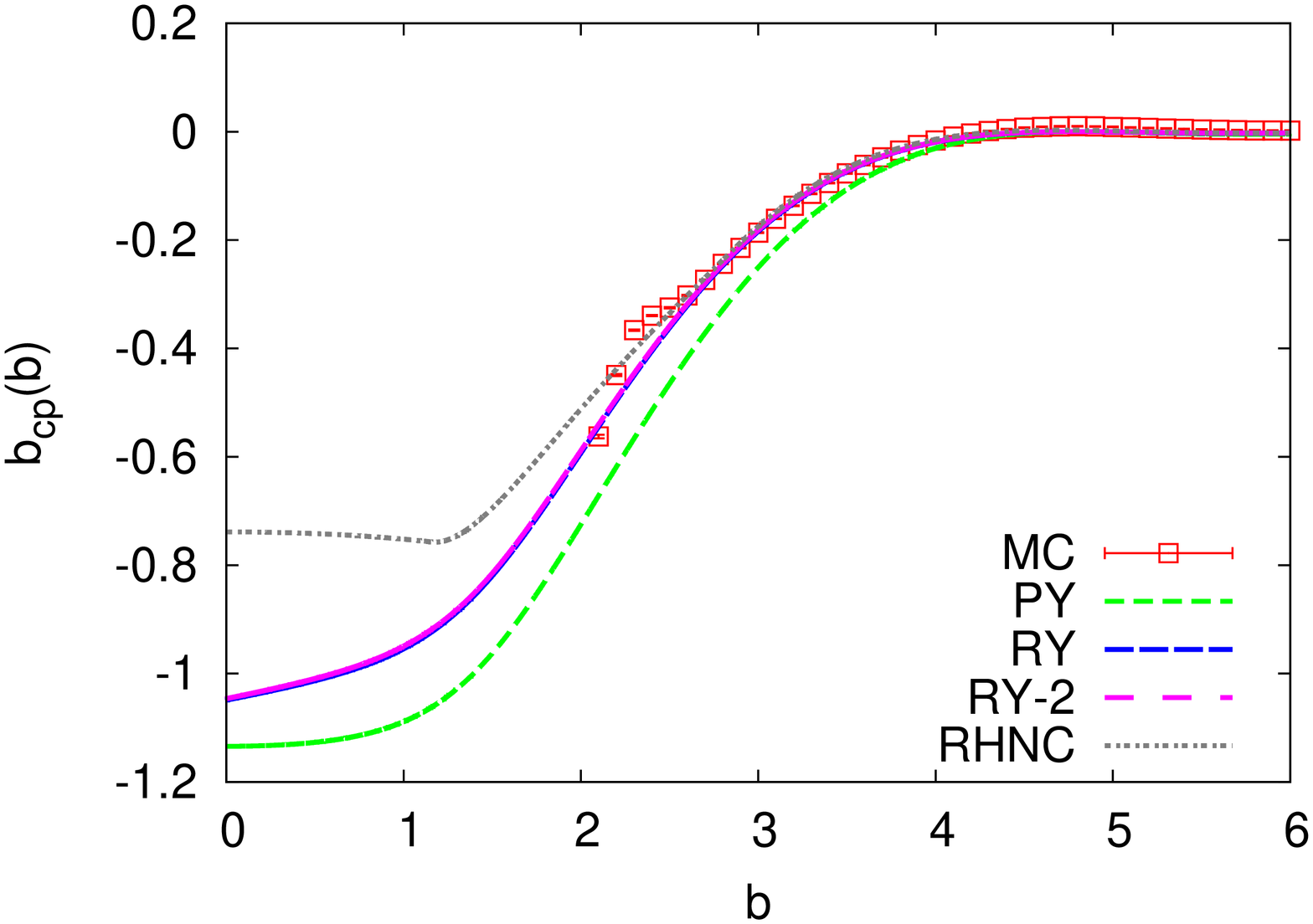,angle=0,width=4.8truecm,angle=-90} 
   \hspace{0truecm} &
\epsfig{file=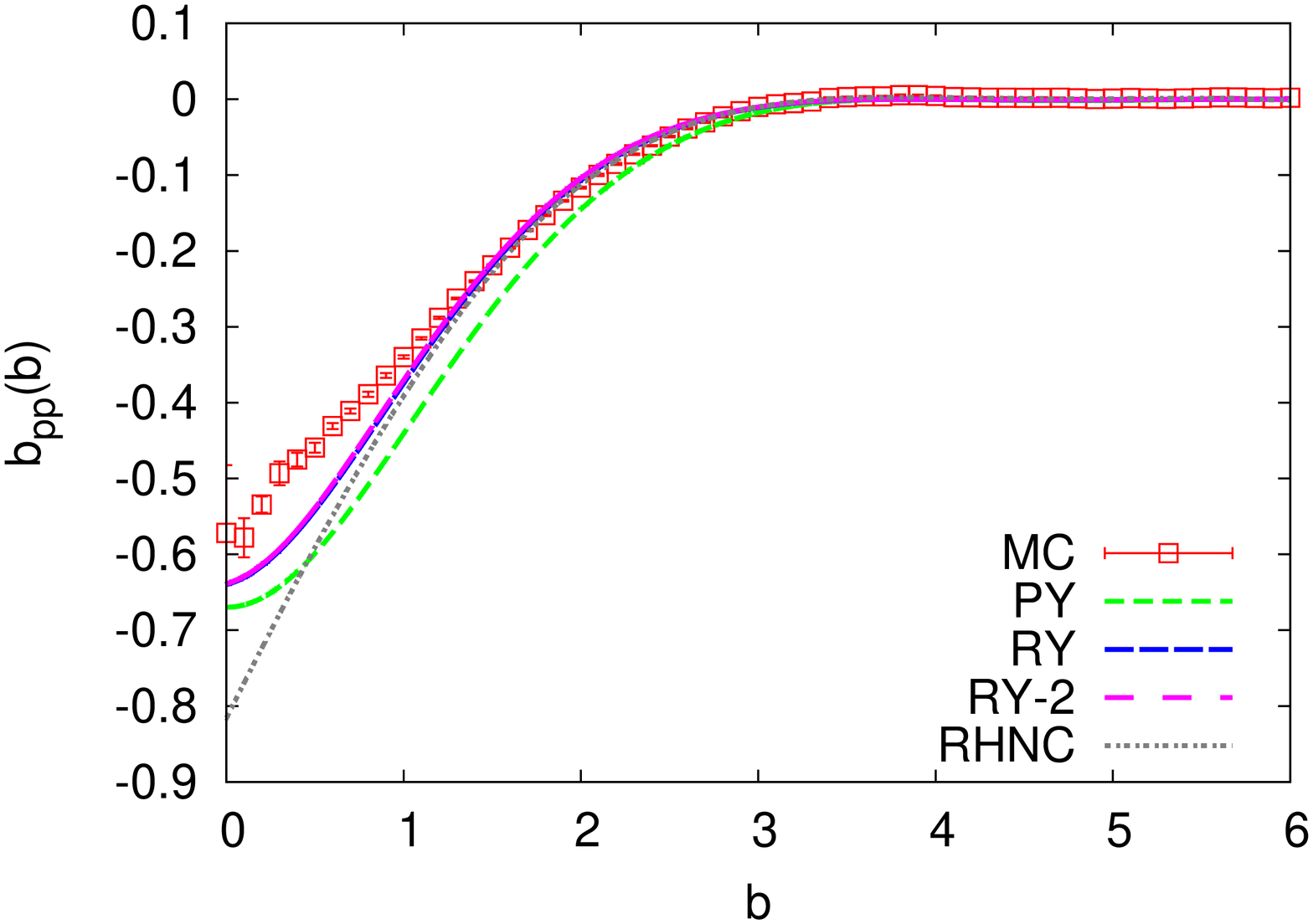,angle=0,width=4.8truecm,angle=-90}  \\
\end{tabular}
\end{center}
\caption{Bridge function for $q=0.5$ as a function of $b = r/R_g$: 
on the left we report 
$b_{cp}(r)$, on the right $b_{pp}(r)$. Top: $\Phi_c = 0.1$;
bottom: $\Phi_c = 0.3$.
We report the Monte Carlo estimates (MC) 
as well as those obtained by using the different closures. 
RY-2 labels the results obtained by using the two-parameter
RY closure discussed in the text. 
}
\label{bridgeq0.5}
\end{figure}

The failure of integral-equation methods to reproduce the 
thermodynamics for $\Phi_c \gtrsim 0.2$ and to provide a 
reasonably accurate estimate of the boundary of the two-phase region
clearly indicates that none of the closures we used is appropriate for the 
problem at hand. To understand better the origin of the discrepancies, we 
now compare the bridge functions used in the integral-equation approach
with the exact estimates obtained numerically, by using the MC results 
for the pair correlation functions. For this purpose we should compute 
$g_{\alpha\beta}(r)$ accurately on large boxes. It turns out that 
this is feasible only for $\Phi_p\to 0$, the case we will study below. 

The input numerical quantities are $g_{cc}(r)$ (we use the accurate 
expressions that can be obtained as 
discussed in Refs.~\cite{MCSL-71,ELLAL-84}), 
$g_{cp}(r)$, and $g_{pp}(r)$. To determine
the last two quantities, we perform simulations for different
values of $\Phi_p$ on systems of linear size $L/R_g = 32$, 24 for 
$q = 0.5$ and 1,  and perform 
an extrapolation to $\Phi_p \to 0$. Then, we determine the direct 
correlation functions by inverting the OZ relations, which,
for $\Phi_p\to 0$, simplify to
\begin{eqnarray}
\hat{c}_{cc}(k) &=& {\hat{h}_{cc}(k)\over 1 + \rho_c \hat{h}_{cc}(k)} ,
\nonumber \\
\hat{c}_{cp}(k) &=& \hat{h}_{cp}(k) - \rho_c \hat{c}_{cc}(k) \hat{h}_{cp}(k),
\nonumber \\
\hat{c}_{pp}(k) &=& \hat{h}_{pp}(k) - \rho_c \hat{c}_{cp}(k) \hat{h}_{cp}(k).
\label{OZ-Phip0}
\end{eqnarray}
Finally, we define
\begin{equation}
b_{\alpha\beta}(r) = \ln\left[ g_{\alpha\beta}(r) e^{\beta V_{\alpha\beta}(r)}
    \right] + c_{\alpha\beta}(r) - h_{\alpha\beta}(r).
\label{bridge}
\end{equation}
We will focus on the polymer-polymer and colloid-polymer functions, 
as $b_{cc}(r)$ depends only on the hard-sphere fluid, a case that has 
already been extensively discussed in the literature. Note that 
$\beta V_{cp}(r)$ is large for $r \lesssim R_c$, so that $g_{\alpha\beta}(r)$
is not determined accurately for these distances. Hence, we are not able 
to obtain reliable estimates of $b_{cp}(r)$ for $r \lesssim R_c$.

For the HNC or the HNC/PY closure, we have $b_{cp}(r) = b_{pp}(r) = 0$. 
In all other cases, the bridge functions are obtained from
Eq.~(\ref{bridge}), using the correlation functions obtained by means of 
the different closures. For the values of $r$ for which $V_{cp}(r)$ is 
large, it is convenient to express $g_{cp}(r) e^{\beta V_{cp}(r)}$ 
in terms of $h_{cp}(r)-c_{cp}(r)$ using the closure relation.
This trick allows us to compute the bridge functions $b_{cp}(r)$ 
inside the core region $r \lesssim R_c$, although here they cannot be 
compared with the Monte Carlo results.
In this section we do not consider the HNC/PY, as 
it has the same bridge functions of the HNC closure. We will instead  
discuss the full PY closure, in which Eq.~(\ref{PY-closure}) is used for all
correlations.  

The bridge functions for $\Phi_c = 0.1$ and $0.3$ are reported in 
Figs.~\ref{bridgeq1} and \ref{bridgeq0.5} for $q=1$ and 0.5, respectively. 
For $\Phi_c = 0.1$ the bridge functions are tiny, explaining why the HNC 
closure works reasonably well. 
The PY and RY closures are essentially equivalent. 
Small deviations are 
evident for $q = 1$ and $r \lesssim 2 R_g$ --- but in this range data 
become increasingly less accurate --- while for $q= 0.5$ no deviations
are observed in the region in which data appear to be reliable.
As $\Phi_c$ 
increases, the bridge functions become increasingly negative for small values 
of $r$. For $q = 1$ and $\Phi_c = 0.3$, none of the closures appear to be 
accurate, although the RY closure is marginally better, and large 
deviations are observed for $r \lesssim 2 R_g$. For $q = 0.5$ the RY 
closure reproduces well $b_{cp}(r)$ up to $r \approx 2 R_g$---the region
outside the colloid core. On the other hand, deviations are clearly
observed for $b_{pp}(r)$ when $r \lesssim R_g$. The PY closure is clearly 
worse, as it underestimates both bridge functions for 
$r \lesssim 2 R_g$-$3 R_g$.

The RY optimization at $\Phi_p = 0$ uses only the colloid-colloid correlations.
Indeed, in this limit the consistency condition is 
\begin{eqnarray}
&& \left({\partial \beta P^{({\rm vir})} \over \partial \rho_c} 
    \right)_{\rho_p = 0} = 1 - \rho_c \hat{c}_{cc}(0).
\label{RY-consistency-1}
\end{eqnarray}
Therefore,
one might think that the relatively poor agreement for the polymer-polymer
correlations for small values of $r$ 
is related to the fact that the procedure does not take into
account polymer properties. We have thus considered a two-parameter 
optimization. We set $\chi_{pp}=\chi_1/R_g$ and $\chi_{cc}=\chi_2/R_c$ 
as free parameters,
while $\chi_{pc}$ is, somewhat arbitrarily, set equal to 
$(\chi_1 + \chi_2)/(R_g + R_c)$.
As consistency conditions, we consider Eq.~(\ref{RY-consistency-1})
and \cite{BenNaim}
\begin{eqnarray}
&& \left({\partial \beta P^{({\rm vir})} \over \partial \rho_p} 
    \right)_{\rho_c,\rho_p = 0} = 1 - \rho_c \hat{c}_{cp}(0),
\end{eqnarray}
which involves polymer-colloid correlations.
In Figs.~\ref{bridgeq1} and \ref{bridgeq0.5},
we also report the bridge functions for this case (they are labelled RY-2). 
For $q = 1$ we observe a significant improvement with respect to the 
one-parameter RY case, although significant differences with Monte Carlo data
are still present for $r/R_g\lesssim 1$. 
For $q = 0.5$ instead, 
the two different RY closures yield equivalent estimates.

As a final case, we consider the RHNC closure, which relies on the 
assumption that the bridge functions can be accurately parametrized by those
of a binary additive hard-sphere mixture. To verify if this is the case, 
we consider $q = 0.5$ and $\Phi_c = 0.3$, and
compute 
\begin{equation}
\Delta(R_p) = \int \left| b_{pp}^{MC}(r) - b_{pp}^{HS}(r,R_p) \right| \, r^2 dr,
\end{equation}
for different values of the effective polymer radius $R_p$. 
The optimal value (minimal $\Delta$) is obtained for $R_p = 0.842 {R}_g$.
We can compare this result with that obtained by using the 
Lado criterion \cite{Lado-82,ELLAL-84}.
For $\Phi_p = 0$, Eq.~(\ref{Lado-eq}) is satisfied as we use
the very accurate hard-sphere correlation function of Ref.~\cite{GH-72}.
To determine $R_p$ one needs to consider the linear term in the polymer density,
i.e., the equation
\begin{equation}
\int r^2 [h_{cp}(r) - h_{cp}^{HS}(r;R_p,R_c)] 
   {\partial b_{cp}^{HS}(r;R_p,R_c)\over \partial R_p} = 0.
\end{equation}
Alternatively, one can 
determine $R_p$ for several small values of $\Phi_p$, performing at the 
end an extrapolation to $\Phi_p\to 0$.
The first method gives $R_p = 0.837 {R}_g$, while the second one
gives $R_p = 0.828 R_g$. Both results are
very close to the 
estimate $R_p = 0.842 {R}_g$ obtained by a direct matching of the bridge
functions.
This confirms that the Lado criterion provides the bridge functions
that are the best approximations of the exact ones.
The resulting bridge 
functions are reported in 
Fig.~\ref{bridgeq0.5}. The RHNC estimate of $b_{cp}(r)$ is in 
agreement with the Monte Carlo function for $r \gtrsim 2 R_g$.
As for $b_{pp}(r)$, 
the RHNC estimate agrees with the Monte Carlo one for 
$r \gtrsim R_g$. At smaller distances, instead, the RHNC bridge function
underestimates the correct one and appears to 
provide a worse approximation than the RY closure.

This analysis for $\Phi_p = 0$ further confirms the results obtained in 
Sec.~\ref{sec3.2}. For $\Phi_c = 0.1$, the bridge functions are 
quantitatively small, confirming the accuracy of the HNC approximation.
On the other hand, for $\Phi_c = 0.3$,
the RY closure is the one that provides the best
approximation, while the HNC closure is the less accurate one as 
it cannot reproduce the small-distance behavior of the bridge functions. 
Note that, while $b_{cp}(r)$ is correctly reproduced in the relevant region
$r \gtrsim R_c$, the polymer-polymer bridge function is always poorly
reproduced  for $ r \lesssim R_g$. This discrepancy gives
rise to similar discrepancies in the correlation functions, as 
discussed in Sec.~\ref{sec3.2}.

\subsection{Integral equations with Monte Carlo bridge functions} 
\label{sec3.4}

\begin{figure}[t]
\begin{center}
\begin{tabular}{cc}
\epsfig{file=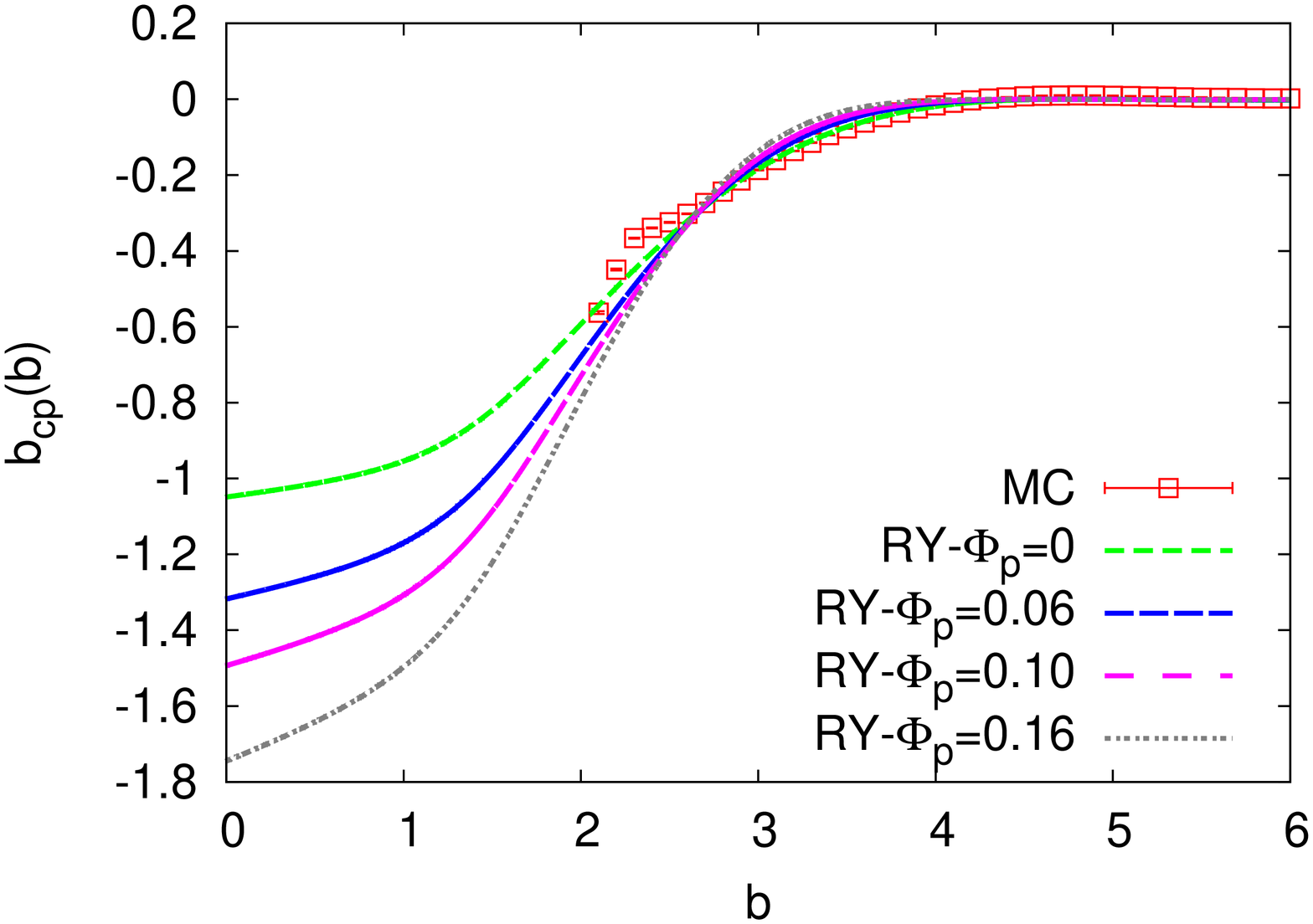,angle=0,width=4.8truecm,angle=-90} 
   \hspace{0truecm} &
\epsfig{file=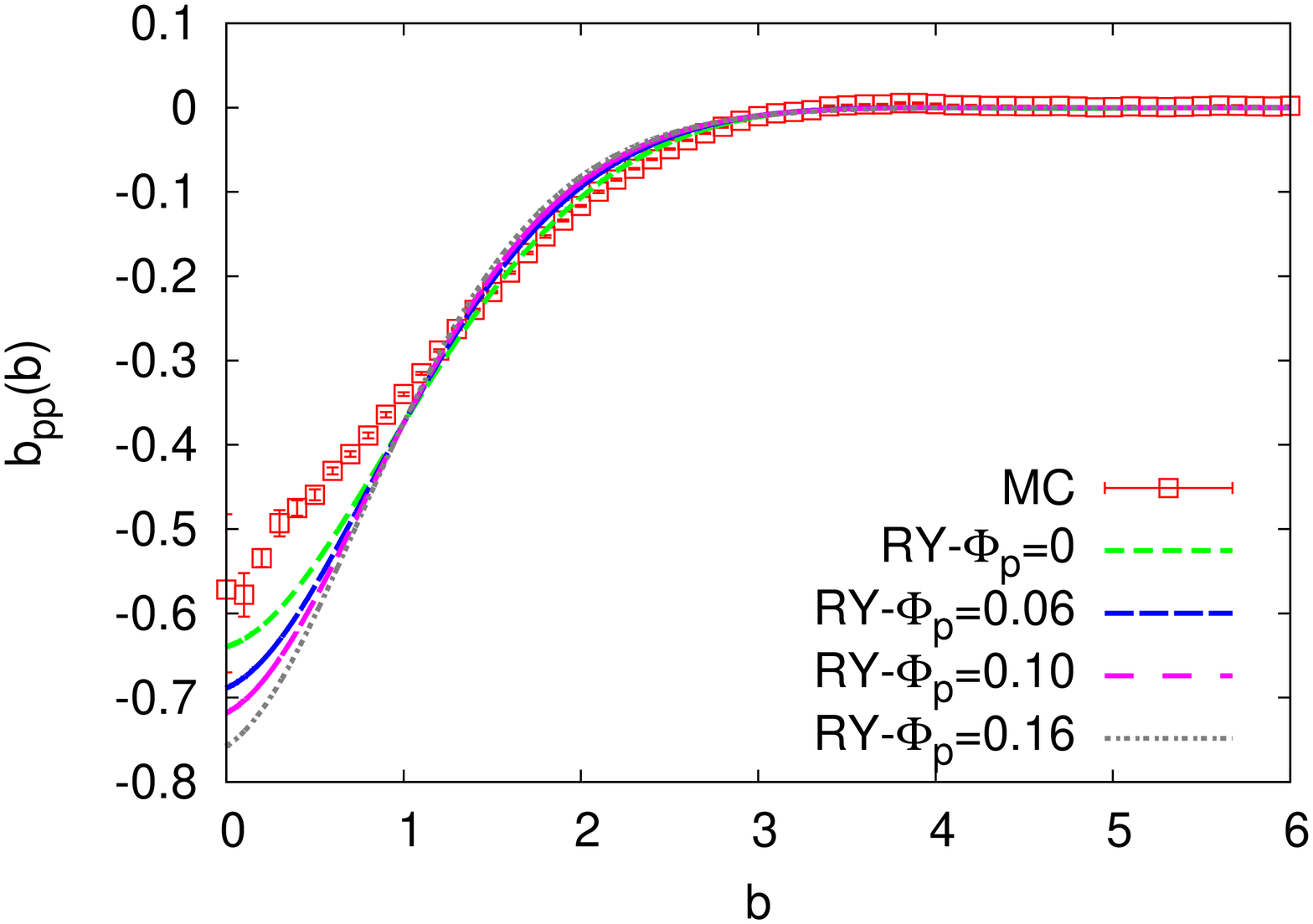,angle=0,width=4.8truecm,angle=-90}  \\
\end{tabular}
\end{center}
\caption{Bridge functions for $\Phi_c = 0.3$ and $q = 0.5$ as 
a function of $b = r/R_g$.
We report the zero-density function obtained by Monte Carlo simulations (MC),
and the RY functions for different values of the polymer volume fraction
$\Phi_p$.
}
\label{bridgeRY-Phip}
\end{figure}

As a final test we decided to determine the solutions of the 
integral equations by using the zero-density Monte Carlo bridge 
functions computed in Sec.~\ref{sec3.3}. In other words, we 
consider the closure relation (\ref{def-bridge}), setting
for all values of $\Phi_p$, 
$b_{pp}(r;\Phi_c,\Phi_p) = b_{pp}^{MC}(r;\Phi_c,\Phi_p=0)$,
$b_{cp}(r;\Phi_c,\Phi_p) = b_{cp}^{MC}(r;\Phi_c,\Phi_p=0)$,
and 
$b_{cc}(r;\Phi_c,\Phi_p) = b_{cc}^{HS}(r;\Phi_c)$, 
where the last quantity is the bridge function of a pure 
hard-sphere system \cite{HG-75}. This approximation is exact for 
$\Phi_p = 0$ and one may wonder whether it provides a reasonable 
approximation also for $\Phi_p > 0$. We have tested the approach for $q=0.5$
and $\Phi_c = 0.3$. The results for the structure factors, 
reported in Fig.~\ref{Scq0.5} (they are labelled MC-B), show that this 
approach is only marginally better than that based on the HNC closure. 
Also the termination point, $\Phi_p = 0.11$, is only slighly above the 
HNC one, $\Phi_p = 0.090$.

To clarify the origin of the discrepancies,
we have determined the RY bridge functions for several values of $\Phi_p$.
As the RY estimates reasonably agree with the Monte Carlo data up to the 
termination line, we take them as estimates of the exact 
density-dependent $b_{\alpha\beta}(r;\Phi_c,\Phi_p)$. As one can see from
the results shown in Fig.~\ref{bridgeRY-Phip}, the density dependence 
of the bridge functions is not large (for $b_{cp}(r)$ the relevant 
region is $b = r/R_g\gtrsim 2$). Yet, this relatively small difference 
is the cause of the different results obtained. In practice, this simple
exercise shows that results are extremely sensitive to the specific form 
of the bridge functions in the colloid-liquid phase $\Phi_c \gtrsim 0.25$.
Hence, accurate results can only be obtained by using accurate bridge 
functions, that none of the methods we investigated is able to provide.

\section{Conclusions} \label{sec4}

In the last years there has been a widespread interest in soft-matter
systems characterized by the presence of macromolecules of mesoscopic 
size. In many situations, if one is only interested in the thermodynamic 
behavior or in structural properties on scales much larger than 
atomic distances, one can use coarse-grained (CG) models in which each 
macromolecule is represented by a single effective particle
\cite{Likos-01,HL-02,DMPP-15}. At variance with simple fluids for 
which potentials always have a hard core, in CG models 
potentials may be soft, allowing different effective molecules to overlap
with a little energy penalty. Monocomponent CG models
have been extensively studied \cite{Likos-01,HL-02} by a variety of techniques. 
Among them, integral-equation methods have been proved to be very 
accurate. In particular, because of the soft nature of the interactions,
the HNC and RY closures work quite well \cite{LBHM-00,WLL-98}. It is then
natural to investigate whether integral equations can be successfully 
applied to the study of the phase diagram and thermodynamics of more 
complex systems, for instance to mixtures of 
macromolecules and colloids, characterized by the simultaneous presence 
of soft and hard-core potentials.

In this paper, we consider a particular CG model, appropriate 
to describe long linear polymers interacting with hard-sphere colloids under 
good-solvent conditions, a well-studied paradigmatic model whose 
phase behavior has been extensively studied, see, e.g., 
Refs.~\cite{LT-11,DPP-14-GFVT}. However, the conclusions should have general 
validity, applying to generic systems with soft and hard-core potentials. 
The phase diagram of the CG model has been
discussed recently in Ref.~\cite{DMPP-15}. The binodal curves and the critical 
points were determined for $q = 0.5$ and $q=0.8$, while, somewhat surprisingly,
no sign of phase separation was found for $q=1$ up to relatively large polymer
densities. Here, we have compared the Monte Carlo results  with predictions
obtained by using integral-equation methods and a variety of different 
closures: HNC, HNC/PY, RY, and RHNC. 

For small values of $\Phi_c$ we find that HNC is quite succesfull in 
predicting the correct thermodynamics and structure. On the other hand,
for $\Phi_c = 0.3$ (note that the critical point of the fluid-fluid transition
is located at $\Phi_{c,\rm crit} = 0.25$ for both $q=0.5$ and 0.8) 
integral equations fail to converge well below the binodal line
determined by Monte Carlo simulations. Below the termination line 
the RY closure is the one that fares 
best, reasonably reproducing the zero-momentum structure factors and the 
pair correlation functions. Nonetheless, RY integral equations stop 
converging at $\Phi_p = 0.18$, 0.39 for $\Phi_c = 0.3$ and $q=0.5$, 0.8, 
respectively, while the binodal 
is located at significantly larger polymer
densities, at $\Phi_p = 0.38$, 0.75 for the same values of $q$.

The failure of integral equations to provide accurate estimates of the 
phase diagram is probably related to the strong nonadditivity of the 
model. Indeed, similarly large differences are observed in Ref.~\cite{PP-14}
for systems of nonadditive hard-sphere mixtures. 
If the system is asymmetric,
i.e., for $y \lesssim 0.6$ 
($y$ is the ratio of the diameters of the two spheres, a 
quantity which is the analog of $q$), integral equations (and also density
functional theory) are unable to provide quantitatively reliable results for 
the phase diagram. Moreover, discrepancies increase with the amount of 
asymmetry considered.

\bigskip

G.D. acknowledges support from the Italian Ministry of Education
Grant PRIN 2010HXAW77. Computations were performed at the Pisa INFN
Computer Center and at CINECA (ISCRA PHCOPY HP10CFFG8Q project).

\appendix

\section{Technical details} \label{AppA}

In the integral-equation approach, pair and direct correlation functions
are discretized 
on $N$ regularly spaced points, $r_n = n \Delta r$. Moreover, all functions 
are assumed to be zero at a cut-off distance $R_{\rm max} = N \Delta r$. 
Typically, we take $\Delta r = 0.001 R_g$ and $N = 32768$ or 65536.
The grid is extremely fine and reasonably large, to guarantee that 
results are stable with respect to the parameters
$\Delta r$ and $N$.\\  In Table~\ref{table-convergence} we report 
several thermodynamic quantities as a function of $\Delta r$ and $N$
for the HNC closure at
$\Phi_c = 0.3$, $\Phi_p = 0.09$, 
a state point very close to the termination line. 

\begin{table}
\caption{Estimates of the structure factors $S_{\alpha\beta}(k=0)$, 
of the concentration factor $S_c(k)$, of the virial pressure $P^{({\rm vir})}$,
and of the compressibility $\kappa_T$ computed using Eq.~(\ref{chit-c})
for $\Phi_c = 0.3$, $\Phi_p = 0.09$, $q = 0.5$, 
and for the HNC closure. We report
results for several values of $N$ and $\Delta r$.}
\label{table-convergence}
\begin{center}
\begin{tabular}{ c c r r r r r r r}
\hline\hline
 & \multicolumn{3}{c}{$N=32768$} &\hspace{1truemm}
    &\multicolumn{4}{c}{$N=65536$} \\
\cline{2-4} \cline{6-9} 
$\Delta r=$ &$0.001$ & $0.002$ & $0.004$ && $0.0005$ & 
             $0.001$ & $0.002$ & $0.004$\\
\hline
$\beta P^{({\rm vir})} R^3_c$ & 0.956 & 0.955 & 0.953 & &
                                0.956 & 0.956 & 0.955 & 0.953\\
$\beta R^3_c/\kappa_T$ & 1.763 & 1.762 & 1.756 &  &
                                 1.764 & 1.763 & 1.761 & 1.756 \\
$S_{pp}(0)$ & 3.399 & 3.436 & 3.560 &  &
              3.384 & 3.399 & 3.436 & 3.560 \\
$S_{cp}(0)$ & $-$0.792 & $-$0.801 & $-$0.830 &  &
              $-$0.788 & $-$0.792 & $-$0.801 & $-$0.830\\
$S_{cc}(0)$ & 0.263 & 0.264 & 0.271 &  &
              0.261 & 0.262 & 0.264 & 0.271\\
$S_c(0)$ & 0.396 & 0.400 & 0.414 &  &
           0.393 & 0.396 & 0.400 & 0.414\\
\hline\hline
\end{tabular}
\end{center}
\end{table}

Estimates do not change as $N$ changes indicating that the cut-off distance 
is large enough. The step size is more crucial, but $\Delta r =0.001$
should be accurate enough. In the paper, most of the analysis use 
$\Delta r = 0.001$ and $N = 32768$. In a few cases, we have 
checked the results, by changing $\Delta r$ and/or $N$ by a factor of 2.
The independence 
of the results on the chosen parameters allows us to exclude that the 
observed behavior is due either to a too small cut-off distance or to 
a too coarse discretization of the correlation functions.

\section{Pair potentials} \label{AppB}

\begin{table}[!b]
\caption{Coefficients parametrizing $\beta V_{cp}(r;q)$ for different values of 
$q$. The parametrization is accurate 
for $1.91\le b \le 5.38$, $0.90\le b \le 4.54$, and  $0.47\le b \le 4.28$
for $q = 0.5, 0.8, 1$, respectively.
}
\label{tab:coeff-Vcp}
\begin{center}
\begin{tabular}{cccccccc}
\hline
\hline
$q$ & $a_1$ & $e_1$ & $c_1$ & $a_2$ & $e_2$ & $c_2$  & $d_2$ \\
\hline \hline 
0.5	& 0.634486	 & 0.305183      & 2.13936     & 15.1368 & 0.512611 &
1.629090 & 1.30679 \\
0.8	& 0.411558	& 0.318504	&	1.40563 & 13.5385 & 0.728577 &
0.572266 & 1.56655 \\
1.0	& 0.982437	& 0.496784	&	0.98100	& 14.1753 & 0.84914 & 0
& 1.6023262	\\
\hline \hline
\end{tabular}
\end{center}
\end{table}

In this section we report the explicit expressions of the pair potentials.
The model consists of coarse-grained polymers, represented as 
soft particles, and colloids. Polymers interact via 
a pair potential $V_{pp}(b)$ given by \cite{PH-05}
\begin{equation}\label{eq:u2cm}
\beta V_{pp}(b)=\sum_{i=1}^3 a_i \exp(-b^2/c_i^2),
\end{equation}
where $b = r/R_g$, $a_1 = 0.999 225$, $a_2 = 1.1574$, $a_3 = -0.38505$,
$c_1 = 1.24051$, $c_2 = 0.85647$, and $c_3 = 0.551876$. 
Colloids interact as hard spheres: 
\begin{eqnarray}
V_{cc}(r) &= &  0 \qquad r > 2 R_c \nonumber \\
V_{cc}(r) &= &  +\infty \qquad r < 2 R_c .
\end{eqnarray}
The polymer-colloid pair potential depends on $q$. For small
values of $b = r/R_g$, i.e. for $ b < b_{\rm min}$ 
($b_{\rm min} \approx R_c/R_g = 1/q$), the potential $\beta V_{cp}(r;q)$ is
large, hence it is impossible (and practically irrelevant) to estimate it 
accurately. 
For $b \gtrsim b_{\rm min}$
we parametrize it as
\begin{equation}
\beta V_{cp}(r;q)=a_1(q) e^{-[(b-c_1(q))/e_1(q)]^2}+a_2(q)
e^{-[|b-c_2(q)|/e_2(q)]^{d_2(q)}},
\end{equation}
where $b = r/R_g$.
Estimates of the coefficients are reported in 
Table~\ref{tab:coeff-Vcp}. 
To verify the accuracy of the parametrization, we 
compare the estimate of $A_{2,cp} = B_{2,cp}/R_g^3$ ($B_{2,cp}$ 
is the second polymer-colloid virial coefficient) obtained 
by using the parametrized potential and the estimate 
of the same quantity in the full-monomer model \cite{DPP-13-depletion}.
Using the parametrized potentials we obtain 
$A_{2,cp} = 106.79, 41.52, 27.50$ for $q=0.5$, 0.8, and 1, respectively,
to be compared with the full-monomer results 
$A_{2,cp} = 107.4(3)$, 41.7(1), 27.54(6). Differences are small (they 
are less than 0.6\%), confirming the accuracy of the results.

\end{document}